\documentclass{aa}

\usepackage[dvipsnames]{xcolor} 
\usepackage{graphicx, subfigure, amsmath, pifont}
\usepackage{amssymb}
\usepackage{multicol}
\usepackage[breaklinks=true]{hyperref} 
\hypersetup{colorlinks=true,citecolor=Blue}
\usepackage{natbib}
\bibpunct{(}{)}{;}{a}{}{,} 
\usepackage[english]{babel}
\usepackage{blindtext}
\usepackage[normalem]{ulem}

\usepackage{comment}
\usepackage{lmodern}
\usepackage{ulem}

\begin{document} 

 \title{Size and structures of disks around very low mass stars in the Taurus star-forming region}

   \author{
    N.~T.~Kurtovic\inst{1},
    P.~Pinilla\inst{1},
    F.~Long\inst{2},
    M.~Benisty\inst{3, 4},
    C.~F~Manara\inst{5},
    A.~Natta\inst{6},
    I.~Pascucci\inst{7, 8},
    L.~Ricci\inst{9}
    A.~Scholz\inst{10},
    L.~Testi\inst{5, 11}
   }
   \institute{
   Max-Planck-Institut f\"{u}r Astronomie, K\"{o}nigstuhl 17, 69117, Heidelberg, Germany, \email{kurtovic@mpia.de}, \email{pinilla@mpia.de}
   \and Center for Astrophysics $\mid$ Harvard \& Smithsonian, 60 Garden Street, Cambridge, MA 02138, USA.
   \and Unidad Mixta Internacional Franco-Chilena de Astronom\'{i}a, CNRS/INSU UMI 3386, Departamento de Astronom\'ia, Universidad de Chile, Camino El Observatorio 1515, Las Condes, Santiago, Chile.
   \and Univ. Grenoble Alpes, CNRS, IPAG, 38000 Grenoble, France.
   \and European Southern Observatory, Karl-Schwarzschild-Str. 2, 85748 Garching, Germany.
   \and Dublin Institute for Advanced Studies, School of Cosmic Physics, 31 Fitzwilliam Place, Dublin 2, Ireland.
   \and Lunar and Planetary Laboratory, The University of Arizona, Tucson, AZ 85721, USA.
   \and Earths in Other Solar Systems Team, NASA Nexus for Exoplanet System Science.
   \and Department of Physics and Astronomy, California State University Northridge, 18111 Nordhoff Street, Northridge, CA 91130, USA.
   \and SUPA, School of Physics \& Astronomy, University of St. Andrews, North Haugh, St. Andrews KY16 9SS, UK.
   \and INAF-Arcetri, Largo E. Fermi 5, 50125 Firenze, Italy.
    }
   \date{}

 \authorrunning{N.~T.~Kurtovic}
 \titlerunning{Size and Structures of Disks around VLMS in the Taurus}

  \abstract
   {The discovery of giant planets orbiting very low mass stars (VLMS) and the recent observed substructures in disks around VLMS is challenging planet formation models.}
   {We aim to estimate if structures, such as cavities, rings, and gaps, are common in disks around VLMS and to test models of structure formation in these disks.  
    We also aim to compare the radial extent of the  gas and dust emission  in disks around VLMS, which can give us insight about radial drift.}
   {We studied six disks around VLMS in the Taurus star-forming region using ALMA Band 7 ($\sim 340\,$GHz) at a resolution of $\sim0.1''$. The targets were selected because of their high disk dust content in their stellar mass regime.}
   {Our observations resolve the disk dust continuum in all disks. In addition, we detect the $^{12}$CO ($J=3-2$) emission line in all targets and $^{13}$CO ($J=3-2$) in five of the six sources.
   The angular resolution allows the detection of dust substructures in three out of the six disks, which we studied by using UV-modeling. Central cavities are observed in the disks around stars MHO\,6 (M5.0) and CIDA\,1 (M4.5), while we have a tentative detection of a multi-ringed disk around J0433. Single planets of masses $0.1\sim0.4\,M_{\rm{Jup}}$ would be required.
   The other three disks with no observed structures are the most compact and faintest in our sample.
   The emission of $^{12}$CO and $^{13}$CO is more extended than the dust continuum emission in all disks of our sample.
   When using the $^{12}$CO emission to determine the gas disk extension $R_{\rm{gas}}$, the ratio of $R_{\rm{gas}}/R_{\rm{dust}}$ in our sample varies from 2.3 to 6.0, which is consistent with models of radial drift being very efficient around VLMS in the absence of substructures.}
   {Our observations do not exclude giant planet formation on the substructures observed. A comparison of the size and luminosity of VLMS disks with  their counterparts around higher mass stars shows that they follow a similar relation .}


  \titlerunning{Resolved Disks and Structures in Disks around Very Low Mass Stars}
  \maketitle

\section{Introduction}                  \label{sect:intro}

For every  ten stars that are formed in the Milky Way, around two to five brown dwarfs (BDs) also form \citep[e.g.,][]{scholz2012, muzic2019}, and M-dwarfs represent about three-quarters of all the stars in our galaxy. Exoplanet discoveries show that short-period ($<50$\,days) sub-Neptune planets occur more frequently around M-dwarfs than around FGK stars \citep{mulders2015}, but also a few giant planets have been discovered around BDs and very low mass stars \citep[$\lesssim0.1\,M_\odot$, VLMS, e.g.,][]{morales2019}. This implies that planets of a large range of masses can form around these objects, although it remains an open question if these massive objects form as binary companions of the BDs and very low mass stars (VLMS), or as planets. Current models of planet formation through pebble or planetesimal accretion cannot explain the formation of giant planets around VLMS \citep[][]{liu2020}. 

Observations of BDs and VLMS from the near-infrared to centimeter wavelength reveal the existence of circumstellar disks around these objects \cite[e.g.,][]{luhman2006, klein2003, scholz2006, scholz2007, ricci2012, daemgen2016, vanderplas2016, ricci2017b, ricci2017a, sanchis2020}, which are more compact and lower in dust mass when compared to disks around T-Tauri stars  \citep[e.g.,][]{pinilla2017b, wardduong2018, hendler2017, hendler2020}. 
The typical millimeter fluxes of such disks suggest that they have dust masses of few Earth masses, challenging the formation of giant planets through core or pebble accretion \citep{liu2020}.

The core accretion scenario for planet formation assumes collisional growth from sub-$\mu$m-sized dust particles from the interstellar medium (ISM) to kilometer-sized bodies or planetesimals \citep[e.g.][]{pollack1996}. 
The collisions of particles and their dynamics within the disk are regulated by the interaction with the surrounding gas.
Different physical processes lead to collisions of particles and their potential growth, such as Brownian motion, turbulence, dust settling, and radial drift \citep[e.g.,][]{brauer2008}. 
All of these processes have a direct or indirect dependency on the properties of the hosting star, such as the temperature and mass.
For instance, from theoretical calculations, settling and radial drift are expected to be more efficient in disks around VLMS and BDs, with BD disks being 15-20\% flatter and with radial drift velocities being twice as high or even more in these disks compared to T-Tauri disks \citep{mulders2012, pinilla2013}. 

With radial drift being a more pronounced problem in disks around BDs and VLMS, it is still unknown how this barrier of planet formation is overcome in these environments where the disks are more compact, colder, and have a lower mass. Millimeter-sized particles have been detected in BD and VLMS disks through measurements of the spectral index \citep{ricci2014, pinilla2017b}, which are only possible to explain when radial drift is significantly reduced by the presence of strong pressure bumps \citep{pinilla2013}.  
The presence of pressure bumps produces substructures, such as rings, gaps, spiral arms, and lopsided asymmetries, with a different amplitude, contrasts, and locations depending on the origin of the pressure variations \citep[e.g.,][]{pinillayoudin2017, andrews2020}.  
Currently, due to sensitivity limitations, most of our observational knowledge about substructures comes from bright (and probably massive) disks, such as the DSHARP sample \citep{andrews2018}. 
A less biased sample of ALMA observations of disks in the star-formation region of Taurus has demonstrated that at least 33\% of disks host substructures at a resolution of 0.1'', and the disks that do not have any substructures are compact \citep[dust disk radii lower that $\sim$50\,au,][]{long2018, long2019}. 
It remains an open question if compact disks are small because they lack pressure bumps or because current observations lack the resolution to detect rings and gaps in these disks \citep[the scale of a radial pressure bump cannot be smaller than one local scale height, e.g.,][]{dullemond2018}. 

\cite{pinilla2018b} have thus far reported the lowest mass star with a resolved large dust cavity (radius $\sim$20\,au) in the disk around the M4.5 star CIDA\,1. 
In the context of planets creating such a cavity, it has been shown that a high planet-to-stellar mass ratio is needed to open a gap and trap particles in disks around VLMS, because in these cases, the disk scale height at a given location is higher than in moderate or high mass stars \citep{pinilla2017b, sinclair2020}. 
In a typical disk around a VLMS as CIDA\,1, at least a Saturn-mass planet is needed to open a gap in the disk. 
This challenges our current understanding of substructures and the common idea that planets are responsible for their formation, since these disks around VLMS and BDs may not have enough mass to form such massive planets, although they may form from gravitational instability if the disks were much denser in their early stages \citep[e.g.,][]{mercer2020}. 

Based on the previous CIDA\,1 observations, we selected a sample of five disks to observe with ALMA at a resolution of 0.1'' in the Taurus star-forming region, whose properties are similar to CIDA\,1. 
Specifically, these disks around low mass stars are more massive compared to other disks with hosts in the same stellar regime. 
These observations included  $^{12}$CO and $^{13}$CO and aim to estimate how common substructures are around VLMS. 
In addition, as radial drift is expected to be very efficient in these disks, they are excellent laboratories to search for the difference between the radial extent of the  gas and dust in disks, which can be a direct signature of radial drift \citep[e.g.,][]{trapman2019}.

This paper is organized as follows: Section \ref{sect:obs} summarizes the target selection, the ALMA observations, and the calibration of the data. 
Section \ref{sect:results} presents the modeling of the data in the visibility plane for the continuum as well as in the image plane for the $^{12}$CO and $^{13}$CO emission. 
In  Section \ref{sect:discussion}, we discuss our results in the context of different origins for the seen substructures as well as the observed difference between the radial extent of the  gas and dust. 
Section \ref{sect:conclusions} summarizes the conclusions of this paper.

\section{Target selection and observations}                     \label{sect:obs}

\subsection{Target selection}

For the new ALMA observations, we selected five disks around VLMS in the Taurus star-forming region. Given that this is the first small survey of high angular resolution observations of disks around VLMS, the sample was selected to optimize the chances of finding substructures.
The criteria used to select the targets were as follows: (1) the target has been previously observed and detected by either SMA or ALMA in millimeter wavelengths (based on the observations by \citet{andrews2013} and \cite{wardduong2018}, respectively); (2) it has a stellar mass between $\sim0.1-0.2$\,$M_\odot$; and (3) it has a high disk dust mass compared to the stellar mass. The last condition comes from observations of the $M_{\rm{dust}}-M_\star$ relation of transition disks and disks with substructures \citep{pinilla2020}, which shows that those disks usually have higher dust masses compared to others with stellar hosts of similar mass. From the list of targets that fulfilled the conditions, we selected the ones with the lowest optical extinction.

\begin{table*}
\centering
\caption{Source properties used in this work. Spectral Type, $T_{\text{eff}}$, and $L_\odot$ comes from \citet{herczeg2014}. Stellar masses derived following the \citet{pascucci2016} method, using distances inferred from Gaia DR2 \citep{gaia2018}.}
\begin{tabular}{ l|l|c|c|c|c|c } 
  \hline
  \hline
\noalign{\smallskip}
  2MASS Name & Used  & Spectral & $M_\star$   & $T_{\text{eff}}$ & $L_\star$   & d    \\
    & Name       & Type     & [$M_\odot$] & [K]              & [$L_\odot$] & [pc] \\
\noalign{\smallskip}
  \hline
\noalign{\smallskip}
  J04141760+2806096 & CIDA\,1  & M4.5     & 0.19        & 3197        & 0.20        & 135.7 \\

  J04322210+1827426 & MHO\,6& M5.0     & 0.17        & 3125        & 0.06        & 141.9 \\

  J04334465+2615005 & J0433 & M5.2     & 0.15        & 3098        & 0.12        & 173.3 \\

  J04422101+252034  & CIDA\,7  & M5.1     & 0.15        & 3111        & 0.08        & 136.2 \\

  J04202555+2700355 & J0420 & M5.25    & 0.14        & 3091        & 0.07        & 170.4 \\

  J04155799+2746175 & J0415 & M5.2     & 0.15        & 3098        & 0.05        & 135.7 \\
\noalign{\smallskip}  
  \hline
  \hline
\end{tabular}
\label{tab:target_summary}
\end{table*}

\subsection{Observations} \label{sect:obs_reduction}
\begin{figure*}
  \centering
    \includegraphics[width=16.cm]{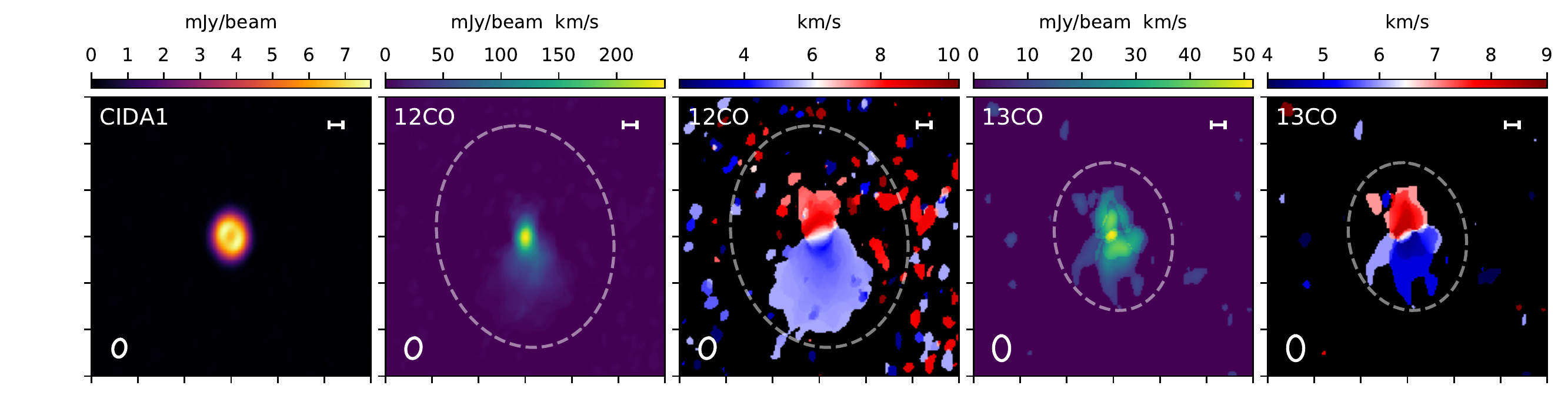}\\  \vspace{-0.15cm}
        \includegraphics[width=16.cm]{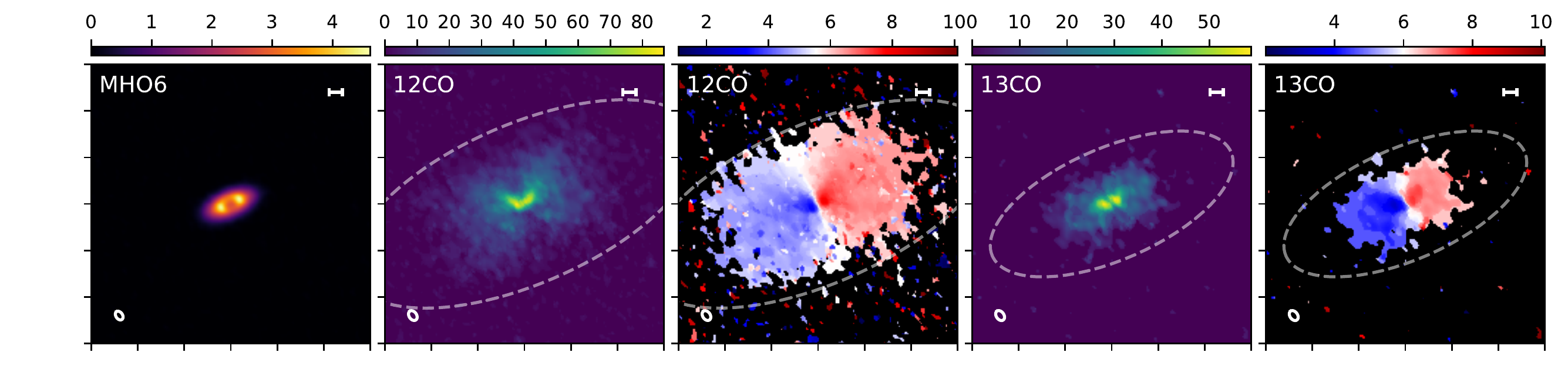} \\  \vspace{-0.15cm}
    \includegraphics[width=16.cm]{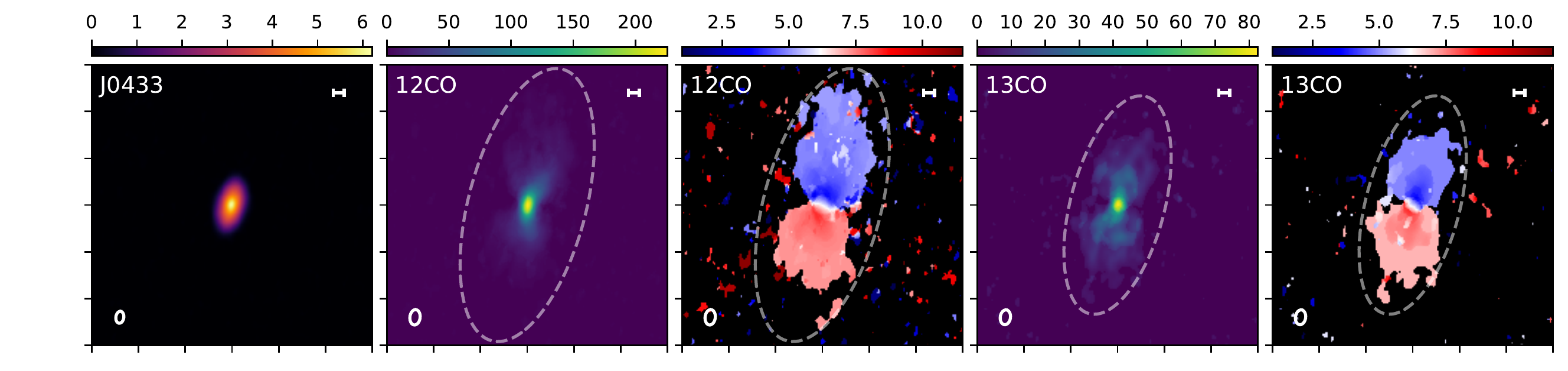}\\  \vspace{-0.15cm}
    \includegraphics[width=16.cm]{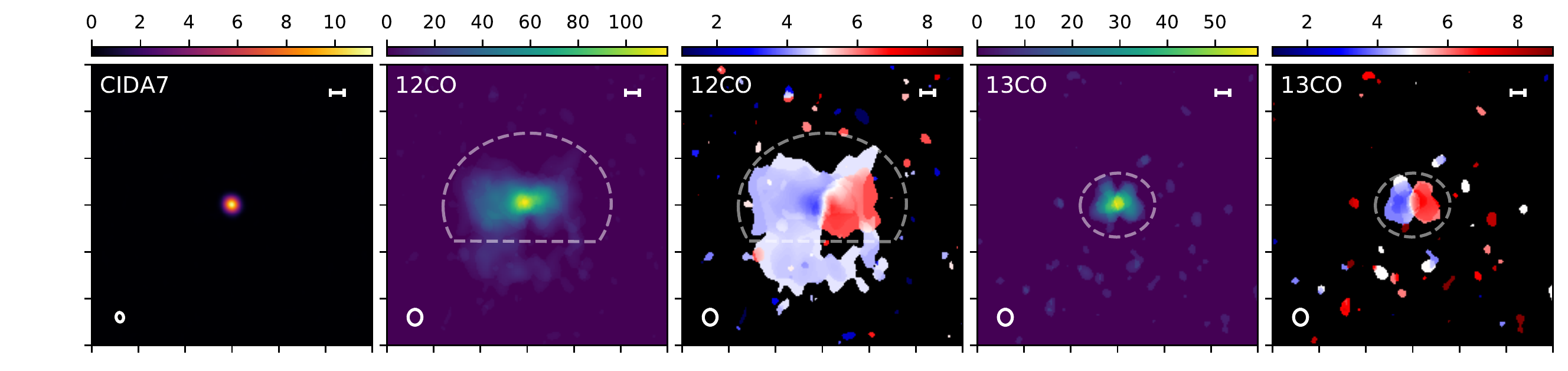}\\  \vspace{-0.15cm}
    \includegraphics[width=16.cm]{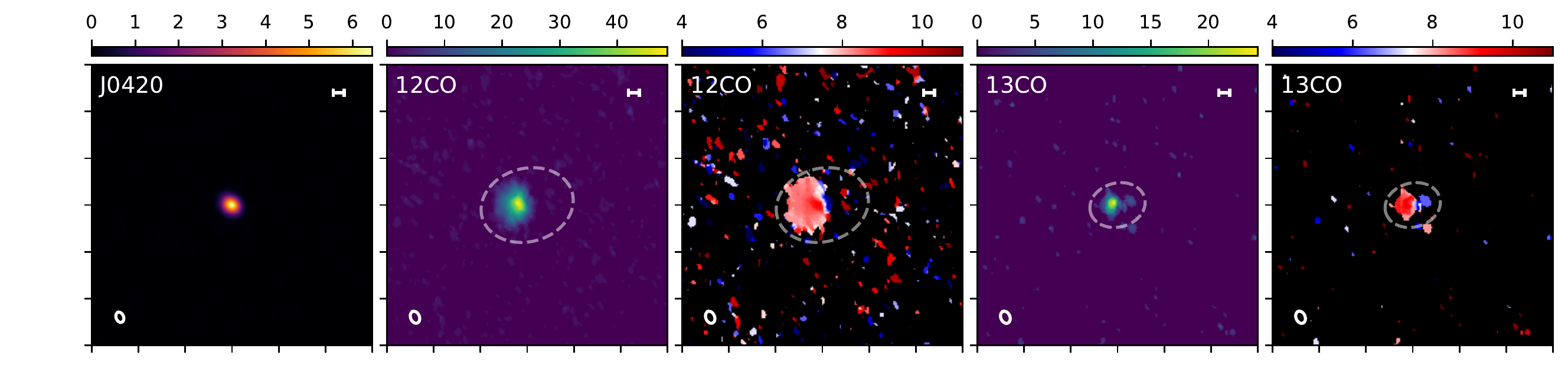}\\  \vspace{-0.15cm}
    \includegraphics[width=16.cm]{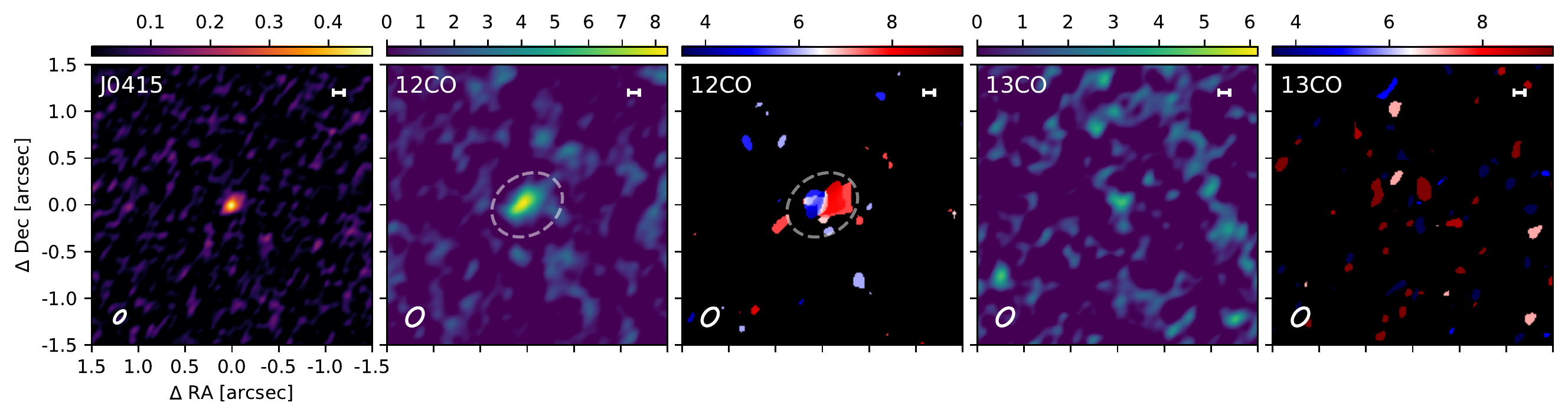}\\  \vspace{-0.15cm}
  \caption{ALMA Observations of the disks around VLMS in Taurus. From left to right: The dust continuum emission, $^{12}$CO moments 0 and 1, and $^{13}$CO moments 0 and 1. All boxes are 3.0'' in size. The scale bar represents 20\,au, and the white ellipses show the synthesized beam. The centers, beam sizes, measured fluxes, and sensitivity are detailed in Tables \ref{tab:imaging_cont} and \ref{tab:imaging_CO}. Dashed lines show the region used to calculate the radial profiles and to measure $R_{68}$ and $R_{90}$.
  A zoomed-in version of the continuum images can be found in Figure \ref{fig:all_uvmodel}.}
  \label{fig:gas_subs}
\end{figure*}

For sample completeness, we included the archival data of CIDA\,1 in our list of targets, given that combining these previous archival observations results in a dataset with a  similar sensitivity and angular resolution. For consistency with future works, the target properties used in this study (shown in Table \ref{tab:target_summary}) were taken from \citet{herczeg2014}, with stellar masses derived following the method described in \citet{pascucci2016}, using distances inferred from Gaia DR2 \citep{gaia2018}. The distances in parsecs was calculated as the inverse of the parallax. The spatial distribution of our sample in the Taurus optical extinction map is shown in Figure \ref{fig:contamination}.

These targets were observed with ALMA at 0.87\, mm (Band~7) as part of the project 2018.1.00310.S (PI: P.~Pinilla), during Cycle 6, with the spectral setup configured to observe in four spectral windows centered at 331.3, 333.3, 344.0, and 345.8\, GHz, with two of them centered to observe dust continuum emission, and two observing molecular line emission from $^{12}$CO\,($J=3-2$) and $^{13}$CO\,($J=3-2$). The frequency resolution for $^{12}$CO was 244.1\, kHz per channel; while for $^{13}$CO and the continuum, it was 976.6\, kHz. The most extended antenna configuration used was C43-8, providing an angular resolution of 0.08'' at best. Some of our sources had archival data from ALMA project 2012.1.00743.S (PI: G. van der Plas) and 2016.1.01511.S (PI: J. Patience), with observations of $^{12}$CO and dust continuum, which were also included in the self-calibration and analysis. Archival Band 7 data of CIDA\,1 was observed by the project 2015.1.00934.S (PI: L. Ricci) published in \citet{pinilla2018b}, and in this work we combined it with 2016.1.01511.S (PI: J. Patience). A summary of the observation details and the data considered for each target are shown in Table \ref{tab:obs_log} in the appendix.

The raw datasets were calibrated by applying the ALMA pipeline using the \texttt{CASA} version specified for each project \citep{casa}. 
Then, we used \texttt{CASA v5.6.2} for the subsequent data handling and imaging. 
We extracted the dust continuum emission of every source by flagging the channels closer than 25\,km\,s$^{-1}$ to the targeted molecular lines. 
To reduce the data volume, we averaged over time (6\,s intervals) and channels (with a width of 125\,MHz). 
Before combining all available observations for each source, the centroid spatial position of the emission was determined by fitting a Gaussian using the \texttt{imfit} task, and shifted using \texttt{fixvis} and \texttt{fixplanets} tasks to the centroid of the observations of  extended baselines, shown in Table~\ref{tab:imaging_cont}. 
We also checked for a consistent flux calibration by comparing the amplitude of the emission in different executions. We found a discrepancy of 12\% in the fourth compact observation of J0433 (2018-11-24), which we rescaled to match all the others. 

In order to boost the signal-to-noise ratio (S/N) of each source, we performed self-calibration of the datasets in two steps: First we combined the compact configuration observations and performed self-calibration. 
Second, we combined those self-calibrated observations with the extended configuration observations and then self-calibrated again. Phase calibrations were applied until the improvement on the S/N was below $\sim$5\%. 
Only 1 amplitude calibration was applied in each step. 
The overall S/N improvement was between $1.5\sim4.0$, depending on the source. The only source where self-calibration was not possible was J0415, because the initial S/N of 9 was too low for improvements. 
The final continuum images were generated using a Briggs robust parameter of 0.0 for CIDA\,1  and 0.5 for the remaining sources. The image properties are summarized in Table \ref{tab:imaging_cont}. 

All the steps for the dust continuum emission calibrations, including centroid shifting, flux calibration, and self-calibration tables, were then applied to the molecular line emission channels. The continuum emission was subtracted using the \texttt{uvcontsub} task, and the images were generated using a robust parameter of 1. In MHO\,6, J0420, and J0415, UV-tappering was applied in order to increase the S/N of the gas images. For MHO\,6, we applied a UV-tappering of 0.13'' on the $^{12}$CO; while for CIDA\,7 and J0415, we applied a UV-tappering of 0.1'' in both molecular line images. All our scripts of self-calibration and imaging are available online \footnote{ \url{https://github.com/nicokurtovic/VLMS_ALMA_2018.1.00310.S} }. 

The final images of the dust continuum, moment 0, and moment 1 of the $^{12}$CO and $^{13}$CO (when detected) are shown in Figure \ref{fig:gas_subs}, and the velocity channel maps are in Appendix \ref{appendix:channel}. The details of the dust continuum and CO images can be found in the appendix as well, summarized in Table \ref{tab:imaging_cont} and \ref{tab:imaging_CO}, respectively.

\section{Results}                       \label{sect:results}

\begin{figure*}
 \centering
        \includegraphics[width=17cm]{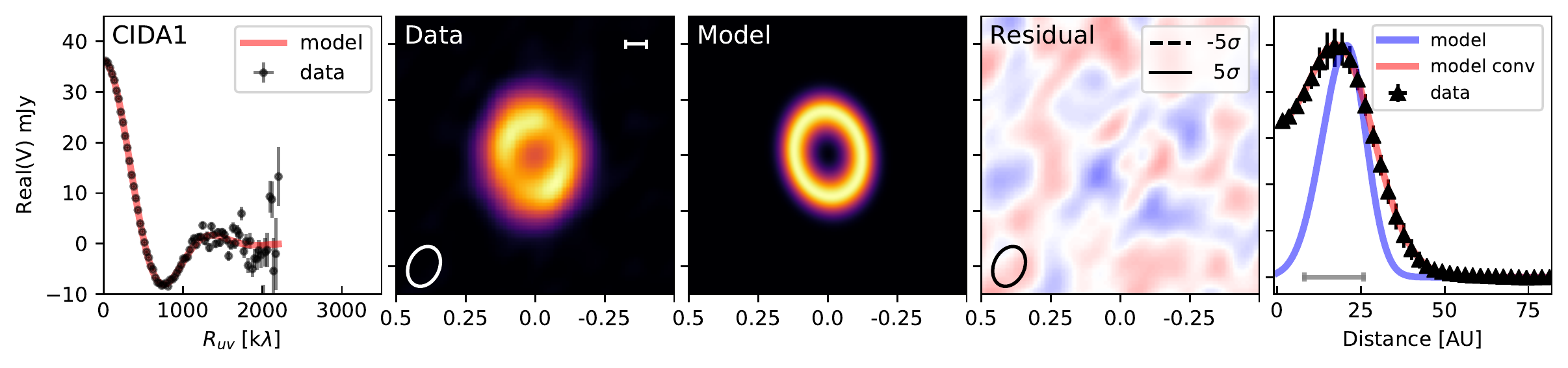}\\ \vspace{-0.2cm}
        \includegraphics[width=17cm]{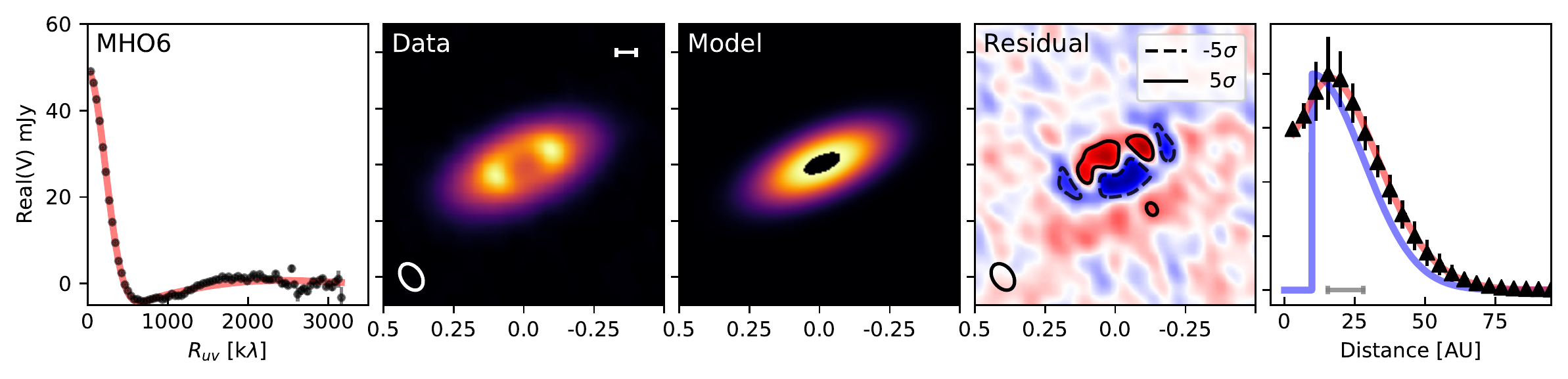} \\ \vspace{-0.2cm}
        \includegraphics[width=17cm]{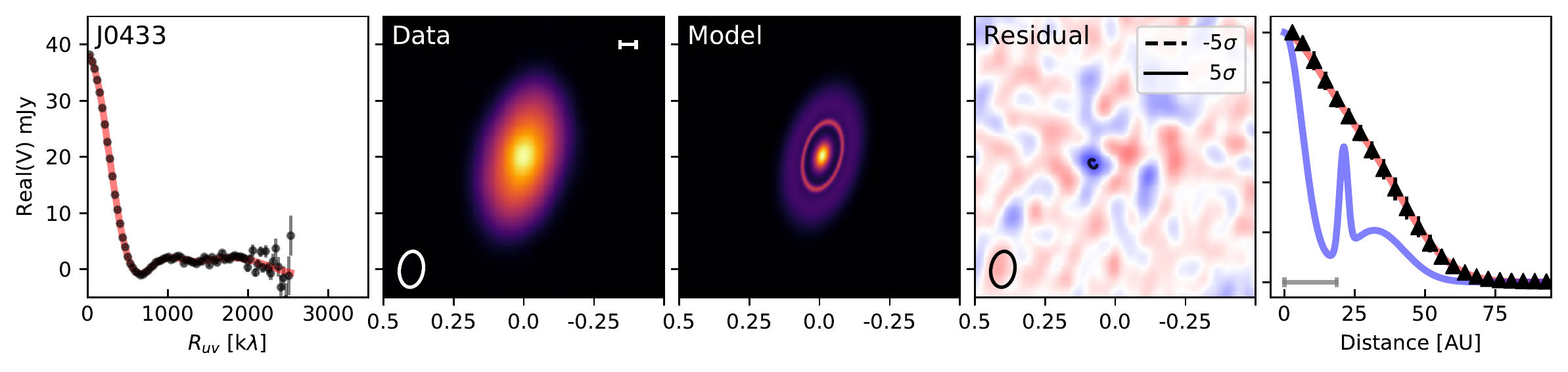}\\ \vspace{-0.2cm}
        \includegraphics[width=17cm]{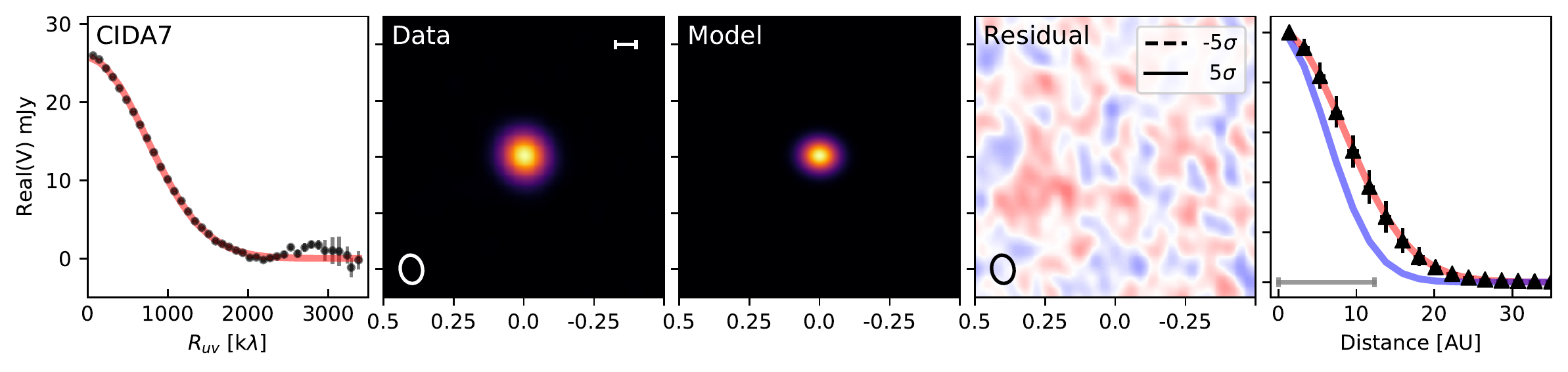}\\ \vspace{-0.2cm}
        \includegraphics[width=17cm]{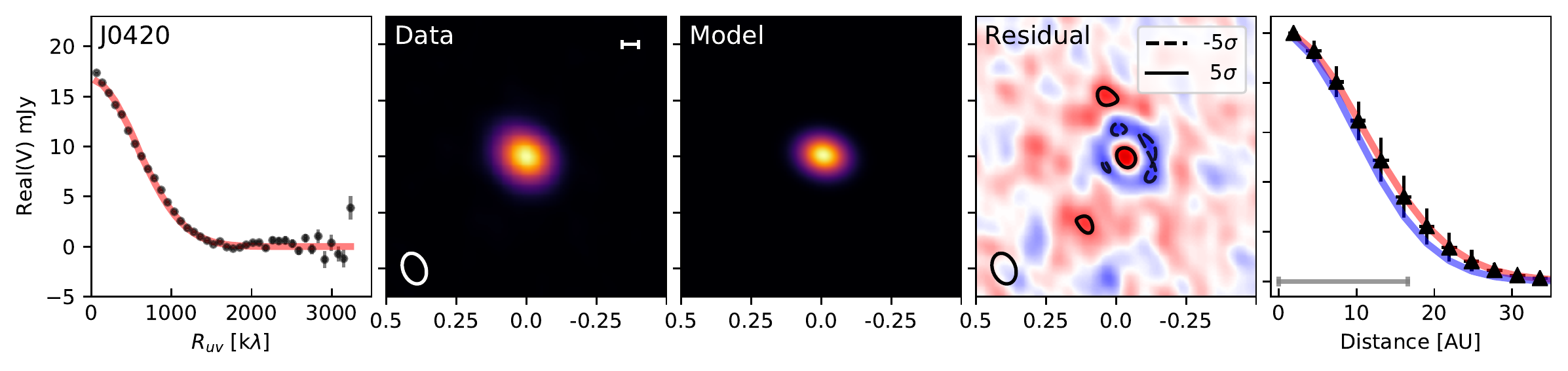}\\ \vspace{-0.2cm}
        \includegraphics[width=17cm]{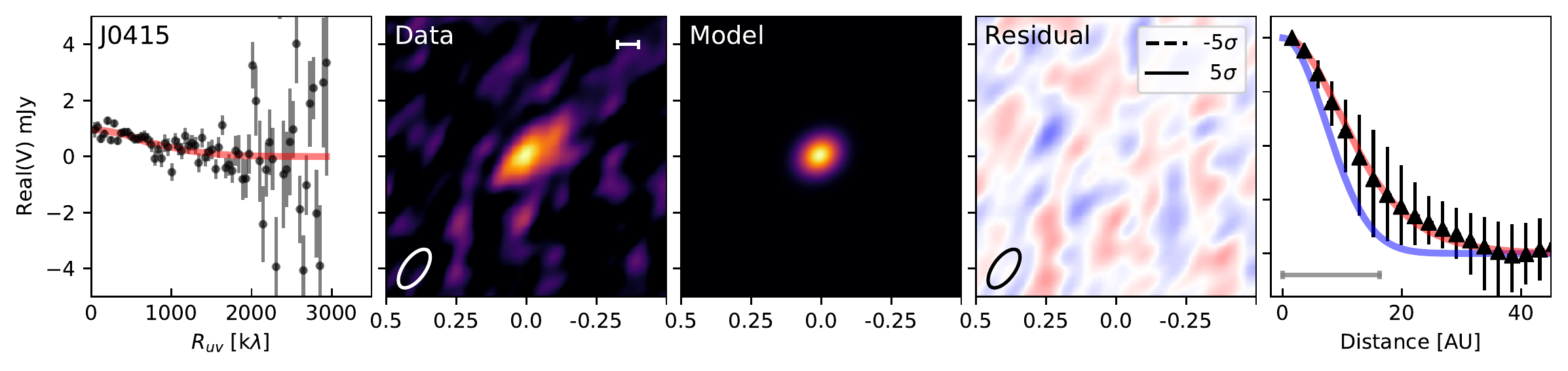}
   \caption{Visibility modeling versus observations of our sample. From left to right: (1) Real part of the visibilities after centering and deprojecting the data versus the best fit model of the continuum data, (2) continuum emission of our sources where the scale bar represents a 10\,au distance, (3) model image, (4) residual map (observations minus model), (5) and normalized, azimuthally-averaged radial profile calculated from the beam convolved images in comparison with the model without convolution (purple solid line) and after convolution (red solid line). In the right most plots, the gray scale shows the beam major axis.}
   \label{fig:all_uvmodel}
\end{figure*}

\subsection{Morphology of the continuum emission}

Our sources are spatially compact (radius of $0.1''\sim0.3''$), and both their radial extent  and substructures have sizes comparable to the synthesized beam shape. To avoid image reconstruction biases, we fit the deprojected brightness profile of the sources in the UV-plane. The continuum visibilities were extracted from the self-calibrated measurement set, and we used the central frequency of each channel to convert the UV-coordinates to wavelength units. We started by modeling every source with a central Gaussian profile, and then we increased the complexity of the profile if the residuals suggested it. We also guided our parametrization of the profiles based on the best fitting from \texttt{frank} \citep{jennings2020}, which fits a nonparametric 1D radial brightness profile in the visibilities, using Gaussian processes. For CIDA\,7, J0420, and J0415, the function that describes their brightness profile is a centrally peaked Gaussian profile, following an intensity given by:

\begin{equation}
    I_{\rm{g}}(r) = f_1 \exp{\left( -\frac{r^2}{2 \sigma_1^2} \right)} \text{,} \label{eq:1g_param_model}
\end{equation}

\noindent where $I_g$ is the Gaussian intensity profile of the source as a function of the radius $r$.

For CIDA\,1  and MHO\,6, we modeled the disk with a radially asymmetric Gaussian ring or a broken Gaussian from hereafter, that is to say the  inner and outer width of the ring  can differ. This profile is motivated by results of radially asymmetric accumulation of particles in pressure bumps \citep[see e.g.,][]{pinilla2015, pinilla2017a}. Such radially broken Gaussian profiles have been used to describe the morphology of different rings in transition disks and disks with substructures \citep[e.g.,][]{pinilla2018a, huang2020}, which is the same model used in \citet{pinilla2018b} to model CIDA\,1.  The intensity profile is given by a ring as follows: 

\begin{equation}
    I_{\rm{bg}}(r) = \left\{
    \begin{aligned}
        f_1 \exp{\left( -\frac{(r - r_1)^2}{2 \sigma_1^2} \right)} \,\,\,\, \text{for }r \leq r_1\\
        f_1 \exp{\left( -\frac{(r - r_1)^2}{2 \sigma_2^2} \right)} \,\,\,\, \text{for }r>r_1
    \end{aligned}  \right. \text{,} \label{eq:bg_param_model}
\end{equation}

\noindent where $I_{\rm{bg}}$ is the broken Gaussian intensity profile as a function of the radius, $r_1$ is the radial location of the ring  peak intensity, and $\sigma_{1,2}$ are the Gaussian widths for the inner and outer sides of the ring, respectively.

Finally, for J0433, the profile is the sum of a centrally peaked Gaussian profile and two symmetric Gaussian rings, as suggested by \texttt{frank}. 
It is also the profile that creates the lowest amount of residuals from our experiments, such as the single Gaussian, Gaussian ring, and broken Gaussian ring.
The intensity profile is 

\begin{equation}
    I_{\rm{J0433}}(r) =  \sum_{i=1}^3  \, f_i \exp{\left(-\frac{(r-r_i)^2}{2 \sigma_i^2} \right)} \text{,} \label{eq:j0433_param_model}
\end{equation}

\noindent where $r_{i=1}=0$, so the first Gaussian is peaked at the center.

The visibilities of each profile were computed by combining each model with a spatial offset ($\delta_{\rm{RA}}$, $\delta_{\rm{Dec}}$), inclination (inc), and position angle (PA). 
Therefore, each model has four more free parameters in addition to those that describe the intensity profile. 
The Fourier transform and the $\chi^2$ calculation were carried out with the \texttt{galario} package \citep{tazzari2018}. 
The pixel size used in the models is 1\,mas, which is several times smaller than the smallest resolvable scale of the observations. 
The $\chi^2$ was scaled up by a factor of 2.667 since \texttt{CASA} does not account for the effective channel width, introduced by Hanning smoothing, when it averages the weights during data binning. 
We adopted a uniform prior probability distribution over a wide parameter range, such that walkers would only be initially restricted by geometric considerations (inc $\in [0,90]$ , PA $\in [0,180]$, $\sigma \geq 0$).

\begin{table*}
\centering
\caption{Best parameters from UV-modeling, following equations (\ref{eq:1g_param_model}), (\ref{eq:bg_param_model}), and (\ref{eq:j0433_param_model}). ``mas'' stands for milliarcsecond. The resulting $F_\text{0.87mm}$ of each model is given in the last row (the measured $F_\text{0.87mm}$ from the data is in Table \ref{tab:imaging_cont}).}
\begin{tabular}{ c|c|c|c|c|c|c|c } 
  \hline
  \hline
\noalign{\smallskip}
                   & CIDA\,1                    & MHO\,6                     & J0433                   &  CIDA\,7                     & J0420                    & J0415                   & unit \\
\noalign{\smallskip}
  \hline
\noalign{\smallskip}
$\delta_{\rm{RA}}$ & $-0.18_{-0.14}^{+0.09}$ & $ 3.84_{-0.09}^{+0.15}$  & $ 3.73_{-0.09}^{+0.07}$ & $-1.18_{-0.06}^{+0.06}$  & $ 2.87_{-0.09}^{+0.08}$ & $-4.19_{-2.68}^{+4.04}$ & mas \\
$\delta_{\rm{Dec}}$& $-5.16_{-0.20}^{+0.11}$ & $-2.69_{-0.07}^{+0.11}$  & $-3.91_{-0.04}^{+0.19}$ & $-0.06_{-0.06}^{+0.06}$  & $-4.42_{-0.10}^{+0.07}$ & $ 0.28_{-4.24}^{+3.32}$ & mas \\
    inc            & $ 38.2_{-0.06}^{+0.15}$ & $64.56_{-0.01}^{+0.06}$  & $57.62_{-0.12}^{+0.01}$ & $31.35_{-0.28}^{+0.30}$  & $38.24_{-0.24}^{+0.26}$ & $34.96_{-28.18}^{+1.21}$& deg \\
    PA             & $ 11.2_{-0.16}^{+0.18}$ &$113.55_{-0.05}^{+0.04}$  &$165.26_{-0.11}^{+0.03}$ & $85.95_{-0.47}^{+0.53}$  & $74.83_{-0.38}^{+0.42}$ &$125.55_{-79.22}^{+22.13}$& deg \\
\noalign{\smallskip}
  \hline
\noalign{\smallskip}
    $f_1$          & $10.25_{-0.01}^{+0.01}$ & $10.45_{-0.01}^{+0.01}$  & $10.76_{-0.01}^{+0.06}$ & $11.00_{-0.01}^{+0.01}$  & $10.63_{-0.01}^{+0.01}$ & $ 9.45_{-0.03}^{+0.08}$ & $\log_{10}$(Jy/sr) \\
    $r_1$          &$153.63_{-1.88}^{+1.69}$ & $68.64_{-0.03}^{+0.17}$  & ---                     & ---                      & ---                     & ---                     & mas \\
    $\sigma_1$     & $54.35_{-1.72}^{+1.69}$ & $0.001_{-0.001}^{+0.05}$ & $36.46_{-4.61}^{+0.43}$ & $45.05_{-0.07}^{+0.08}$  & $57.69_{-0.11}^{+0.11}$ & $53.09_{-8.60}^{+0.45}$ & mas \\
    $f_2$          & ---                     & ---                      & $10.39_{-0.22}^{+0.01}$ & ---                      & ---                     & ---                     & $\log_{10}$(Jy/sr) \\
    $r_2$          & ---                     & ---                      &$121.48_{-5.99}^{+0.58}$ & ---                      & ---                     & ---                     & mas \\
    $\sigma_2$     & $41.92_{-0.97}^{+1.15}$ &$129.20_{-0.08}^{+0.21}$  & $ 8.36_{-0.12}^{+8.59}$ & ---                      & ---                     & ---                     & mas \\
    $f_3$          & ---                     & ---                      & $10.08_{-0.01}^{+0.01}$ & ---                      & ---                     & ---                     & $\log_{10}$(Jy/sr) \\
    $r_3$          & ---                     & ---                      &$184.67_{-3.85}^{+1.77}$ & ---                      & ---                     & ---                     & mas \\
    $\sigma_3$     & ---                     & ---                      & $60.76_{-0.81}^{+1.85}$ & ---                      & ---                     & ---                     & mas \\
\noalign{\smallskip}
  \hline
\noalign{\smallskip}
    $F_\text{0.87mm}$& $36.08 \pm 0.17$      & $49.05 \pm 0.15$         & $37.52 \pm 0.1$          & $25.52 \pm 0.10$        & $16.50 \pm 0.05$        & $0.95 \pm 0.19$         & mJy \\
\noalign{\smallskip}
  \hline
  \hline
\end{tabular}
\label{tab:mcmc_results}  
\end{table*}

We used a Monte Carlo Markov chain (MCMC) routine based on the \texttt{emcee} package \citep{emcee} to sample the posterior probability distribution of each parameter space. Furthermore, we ran more than 250000 steps after converging to find the most likely set of parameters and the error bars, taken from the 16th and 84th percentile.

Our results for each parameter are shown in Table \ref{tab:mcmc_results}, while in Fig.~\ref{fig:all_uvmodel} we show the models with and without convolution (right most panel). The visibilities and radial profile were deprojected using the best inclination and position angle. The residual image was generated in \texttt{CASA} using the same parameters and procedure used for the observations, from a measurement set with its visibilities calculated by subtracting the best model from the data.

\begin{figure*}
 \centering
   \includegraphics[width=16cm]{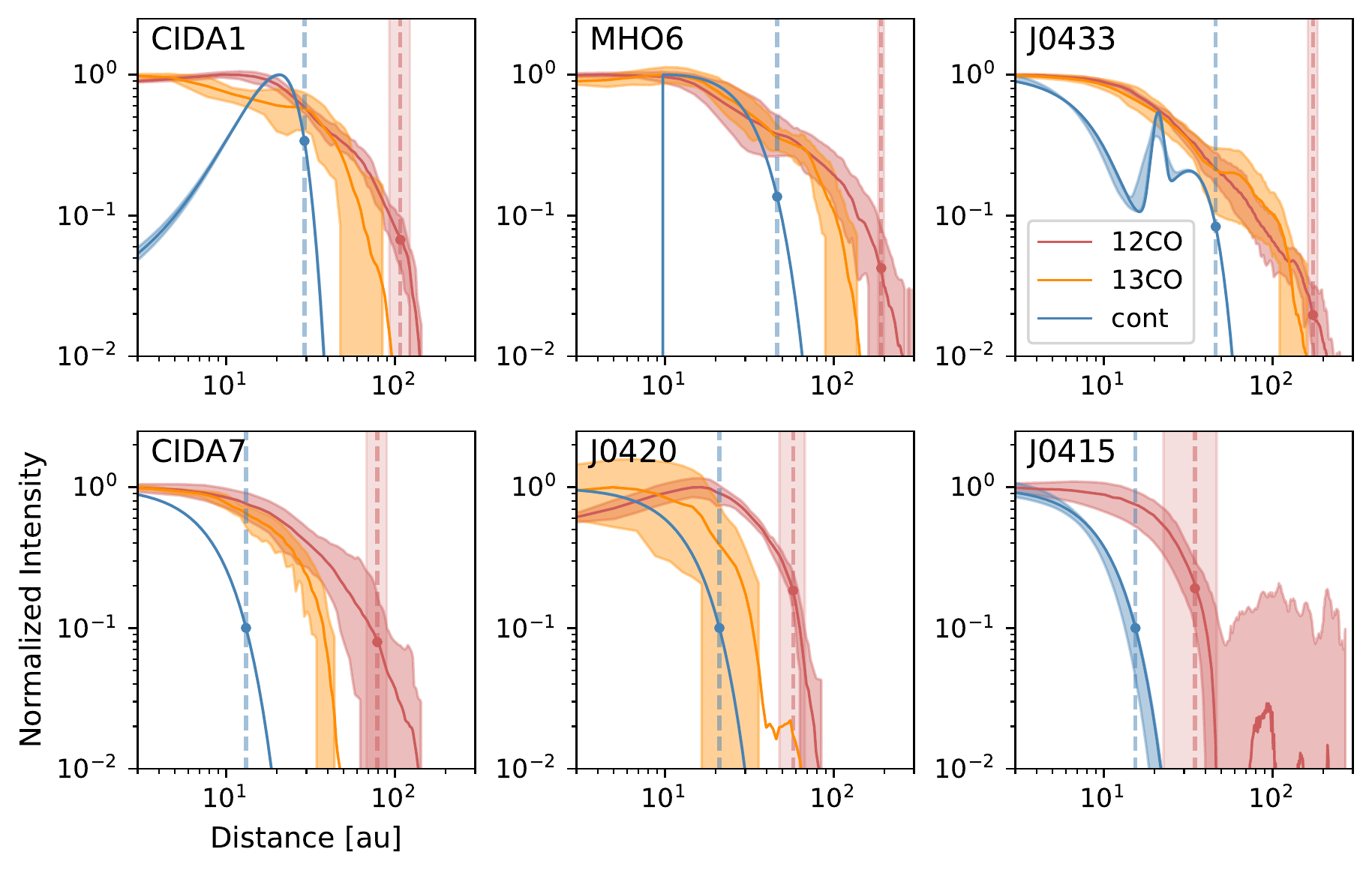}
   \caption{Normalized azimuthally averaged radial profiles of the dust continuum, $^{12}$CO, and $^{13}$CO of each source. Each curve has a 68\% error region. The orange and red solid curves correspond to the average intensity profile for the gas profiles of the deprojected images, while the blue solid line corresponds to the best $\chi^2$ solution from the visibility fit of the dust continuum. The dashed vertical line denotes the position and $1\,\sigma$  error of the $R_{90}$ radius for the dust (blue) and the gas (red).}
   \label{fig:rad_prof_all}
\end{figure*}

The radial profile recovered from the UV-modeling was used to measure the dust continuum emission radii ($R_{\rm{dust}}$) that encloses 68\% and 90\% of the total flux (the dust  $R_{68}$ and $R_{90}$, respectively). 
We used the different sets of parameters sampled by the walkers to compute their radial profiles, and we found the 16th and 84th percentile on the $R_{68}$ and $R_{90}$ to calculate the error bars. 
The results are shown in Table \ref{tab:rgas_rdust} and in Figure \ref{fig:rad_prof_all}.
The continuum emission in our sample extends up to $46\,$au in the biggest disks, MHO\,6, and J0433, while we also have the smallest ($13\,$au) and the dimmest ($\sim1\,$mJy) Taurus disks ever resolved in 0.87\, mm emission, CIDA\,7, and J0415, respectively.

\begin{table*}
\centering 
\caption{Continuum and CO radial extension for each source. Uncertainties for continuum comes from the walkers distribution in each MCMC. Gas radii uncertainties were calculated from the synthesized beam radius of each image. \label{tab:rgas_rdust}}
\begin{tabular}{ c|c|c|c|c|c }
    \hline
    \hline
\noalign{\smallskip}
          &  $R_\% $  & $R_{\rm{dust}}$  [au] & $R_{^{12}\rm{CO}}$  [au]  & $R_{^{13}\rm{CO}}$  [au] & $R_{\rm{gas}} \, /\, R_{\rm{dust}}$   \\
          &       &                        & ($R_\text{gas}$)          &            &   \\
\noalign{\smallskip}
    \hline
    \hline
\noalign{\smallskip}
    CIDA\,1  &  68   & $24.57_{-0.04}^{+0.03}$ &  $73.7\pm15.0$ & $ 47.4\pm 6.8$ & $3.0 \pm 0.6$ \\
          &  90   & $29.22_{-0.09}^{+0.07}$ & $108.1\pm15.0$ & $ 68.5\pm 6.8$ & $3.7 \pm 0.5$ \\

          &       &                         &                &                &               \\
    MHO\,6  &  68   & $34.18_{-0.05}^{+0.01}$ & $130.7\pm 7.0$ & $ 78.5\pm 7.1$ & $3.8 \pm 0.2$ \\
          &  90   & $46.36_{-0.07}^{+0.01}$ & $191.9\pm 7.0$ & $113.0\pm 7.1$ & $4.1 \pm 0.2$ \\

          &       &                         &                &                &               \\
    J0433 &  68   & $36.58_{-0.01}^{+0.11}$ & $114.6\pm11.9$ & $ 92.2\pm11.7$ & $3.1 \pm 0.3$ \\
          &  90   & $46.22_{-0.07}^{+0.08}$ & $174.0\pm11.9$ & $125.5\pm11.7$ & $3.8 \pm 0.3$ \\

          &       &                         &                &                &               \\
    CIDA\,7  &  68   & $ 9.26_{-0.02}^{+0.02}$ & $ 55.1\pm10.8$ & $19.6 \pm10.9$ & $6.0 \pm 1.2$ \\
          &  90   & $13.16_{-0.02}^{+0.02}$ & $ 78.9\pm10.8$ &                & $6.0 \pm 0.8$ \\

          &       &                         &                &                &               \\
    J0420 &  68   & $14.84_{-0.03}^{+0.03}$ & $ 43.1\pm9.9$  &  $23.3\pm9.8$  & $2.9 \pm 0.7$ \\
          &  90   & $21.09_{-0.04}^{+0.04}$ & $ 57.7\pm9.9$  &  $32.2\pm9.8$  & $2.7 \pm 0.5$ \\

          &       &                         &                &                &               \\
    J0415 &  68   & $10.87_{-0.15}^{+1.78}$ & $ 26.4\pm12.1$ & ---            & $2.4 \pm 1.2$ \\
          &  90   & $15.46_{-0.22}^{+2.53}$ & $ 34.8\pm12.1$ & ---            & $2.3 \pm 0.9$ \\

\noalign{\smallskip}
    \hline
    \hline
\end{tabular}
\end{table*}

\subsection{Rings and cavities in CIDA\,1, MHO\,6, and J0433}

We find evidence of ring structures in three out of six disks in our sample. At the current resolution, CIDA\,1  and MHO\,6 show  a single ring and a cavity, which are well described by a broken Gaussian profile. In CIDA\,1, we find the inner side of the Gaussian to be more extended than the outer side, which is similar to the results obtained in \citet{pinilla2018b}. The ratio between the widths of the inner side and the outer side in our analysis is 1.2, and the peak of our ring model is located at 20.8\,au.

For MHO\,6, we find that a strong asymmetric ring peaked at $10\,$au, with its inner side having converged to $\sigma=0\,$au with a narrow error margin. To ensure that our result was not being affected by numerical biases related to the pixel size, we ran another MCMC with a pixel size of 0.4\,mas (about 0.057\,au), and we consistently recovered the same result for each parameter. Even though this steep transition could suggest an unresolved inner side of the ring, the best model is also driven by the non-axisymmetric emission of the disk, which is only noticeable  when looking at the residuals (see Fig.~\ref{fig:all_uvmodel}). The contrast of these asymmetries is about $5\%$ of the peak amplitude of the ring, meaning that most of the emission is still well described by a radially axisymmetric ring.

In J0433, our best model finds two rings located at 21 and 32 \,au, with gaps at 16 and 25 \,au.
The brightness ratio between the first ring and first gap is about 4.4; whereas, the contrast is 1.2 between the second gap and ring. Since the brightness ratio fades at the $1\,\sigma$  error (as seen in Figure \ref{fig:rad_prof_all}), we cannot exclude that it is an extension of the outer side of the first ring.

\subsection{Dust disk masses \label{sec:dust_mass}} 

\begin{figure}
 \centering
   \includegraphics[width=9cm]{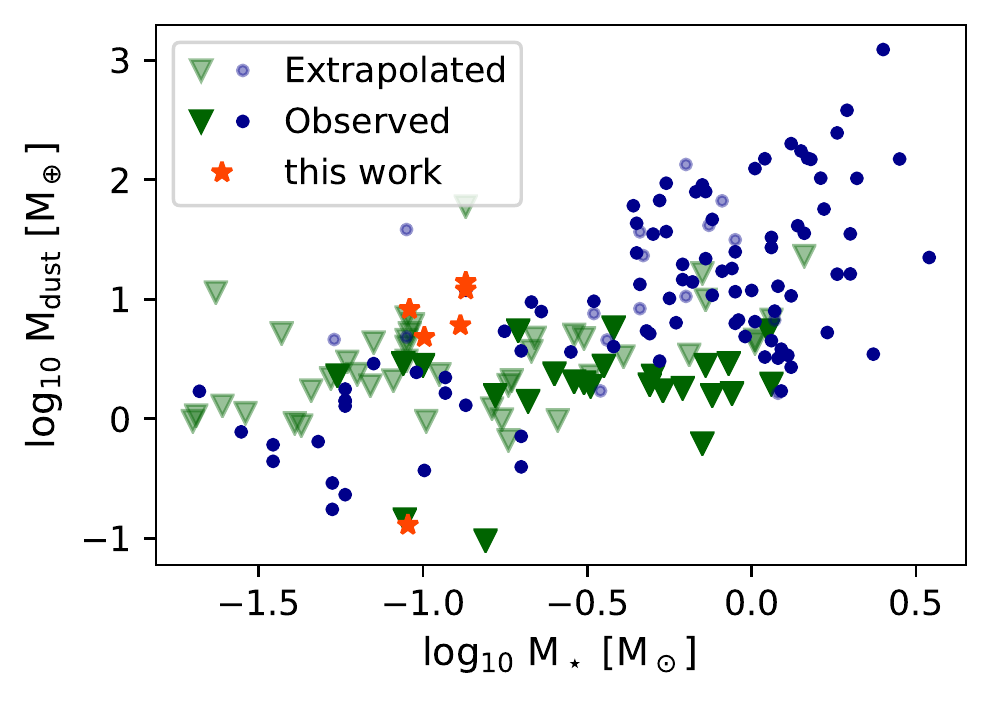}
   \caption{Dust mass compared to stellar mass for the Taurus disks. We note that M$_d$ was computed using distances inferred from Gaia DR2, assuming a 20K disk temperature and using the most updated value of $F_\text{0.87\rm{mm}}$ from \citet{andrews2013}, \citet{tripathi2017}, and \citet{wardduong2018}. Sources where $F_\text{0.87mm}$ was extrapolated from a $F_\text{1.3mm}$ measurement are plotted in faded colors \citep[see Table 2 in][]{andrews2013}, while direct measurements from SMA or ALMA are plotted in darker colors. Green triangles denotes upper limits, and the VLMS studied in this work are plotted as orange stars.}
   \label{fig:mstar_mdust}
\end{figure}

Assuming the dust continuum emission is optically thin, we estimated the disk dust mass by following \citet{hildebrand1983}:
\begin{equation}
    M_{\rm{dust}} \approx \frac{d^2 F_\nu}{\kappa_\nu B_\nu(T)} \text{,}
    \label{eq:mdust}
\end{equation}

\noindent where $d$ is the distance in parsecs (given in Table \ref{tab:target_summary}, taken from Gaia DR2), $\kappa_{\nu}$ is the mass absorption coefficient, and $B_\nu(T)$ is the Planck function. For $\kappa_\nu$, we assumed the opacity law $\kappa_\nu=2.3\,\rm{cm}^2\,\rm{g}^{-1}\,(\nu/230\,\rm{GHz})^{0.4}$ \citep{andrews2013}, while the temperature was assumed constant at $T_{\rm{dust}}=20\,$K \citep[e.g.,][]{ansdell2016, pinilla2018b}. We estimated the uncertainty of the dust mass by taking the $10\%$ uncertainty in the flux calibration. Our results for each source are compiled in Table \ref{tab:imaging_cont} and shown compared with the disk dust mass in the Taurus region in the Figure \ref{fig:mstar_mdust}. The lowest dust mass estimate was obtained for J0415, with only $0.22\pm0.02\,M_\oplus$  (about $2.1\pm0.2\,M_{\rm{Mars}}$), while the most massive dust disk was detected in J0433 with $14.32\pm1.43\,M_\oplus$.

\subsection{$^{12}$CO and $^{13}$CO emission}

\subsubsection{Lines detection in each source}

Both molecular lines were detected in all of our sources, except for J0415, where only $^{12}$CO was observed. As shown in Figures \ref{fig:gas_subs} and \ref{fig:all_uvmodel}, our angular resolution of $\sim90\,$mas was just enough to resolve the dust continuum and gas emission of this source, but the limited sensitivity did not allow us to constrain the geometric parameters as in the other systems.

For the two smooth sources, CIDA\,7  and J0420, we found strong cloud contamination, so that the western side of the $^{12}$CO emission in J0420 is completely absorbed. The rotational pattern was, however, recovered in the $^{13}$CO (see Figures \ref{fig:gas_subs} and \ref{fig:chanmap_j0420}). On the other hand, in CIDA\,7,  we observed an extended asymmetric emission in the south region, with a velocity range of at least 0.8\,km\,s$^{-1}$ (it was detected in two velocity channels). The contribution of this emission is not significant to the total flux of the gas emission, and it was not detected in $^{13}$CO. Although the S/N did not allow us to accurately recover geometric parameters from the gas emission, we can see that for both sources, the PA obtained from the dust continuum emission is in good agreement with the orientation of the major axis in the rotation pattern.

In the sources with detected substructures, both J0433 and CIDA\,1  are affected by cloud contamination.
For CIDA\,1,  the side most affected by extinction is the north side, while the same is true for the south side of J0433 (row 1 and 3 of Figure \ref{fig:gas_subs}).
The least cloud-contaminated source is MHO\,6 disk  (also seen in Fig.~\ref{fig:contamination}), which is the brightest and more extended disk in CO emission of our sample.
Since the cloud contamination is weaker, the gas can be detected continuously in every channel from both high velocity ends. Given our good spatial detection of both lines, we used the package \texttt{eddy} \citep{eddy} to analyze the Keplerian rotation, which we further describe and discuss in Sections \ref{sect:keplerian_rotation} and \ref{sec:mho6kinematics}.

\subsubsection{Radial gas profiles}

The inclinations and position angles obtained from the continuum fitting were used to deproject the distances from the central star in the moment 0 images, and we calculated the azimuthally averaged radial profiles of the $^{12}$CO and $^{13}$CO emission. The vertical structure was neglected in this calculation. In CIDA\,1, J0433, and J0420, the profiles were calculated from the side that is less affected by cloud contamination. In CIDA\,7, we found that the contribution of the asymmetric emission in the south is negligible, but we nevertheless did not take it into account as we wanted to recover the disk emission profile (see masking in Figure \ref{fig:gas_subs}). The gas  $R_{68}$ and $R_{90}$ radii  that encloses 68\% and 90\% of the flux are shown in Table \ref{tab:rgas_rdust} and Fig.~\ref{fig:rad_prof_all}.

As we discuss in Section \ref{sec:differences}, the brightness in dust continuum and gas emission of J0415 are much lower than expected from previous SMA observations \citep{andrews2013}.
The low S/N of the detection prevented us from applying self-calibration, and so the sensitivity is two times worse than in the other VLMS disks.
Therefore, we calculated the radial profile of the $^{12}$CO emission from an image generated by only considering the channels in the velocity range with line detection, and we did not apply clipping at $3\,\sigma$ .

\subsubsection{Keplerian rotation of CO emission} \label{sect:keplerian_rotation}

Although we recovered the rotational pattern for all the disks in our sample in both CO isotopologues (with the exception of the  $^{13}$CO in J0415 ), the strong cloud contamination and low S/N of our images prevented us from reliably recovering the dynamical mass from our sources without proper modeling. 
The only system where the cloud contamination does not completely make the central velocity channels extinct is MHO\,6, where we have good spatial detection of both lines.
As described in Section \ref{sect:obs_reduction}, we used \texttt{CASA} to generate the moment 1 image from the velocity maps; in addition, we generated velocity integrated images using the python package \texttt{bettermoments} \citep{bettermoments}, with a quadratic and Gaussian method, and also by varying the root mean squre (RMS) clipping limit. The Keplerian rotation of these images was modeled using the package \texttt{eddy} \citep{eddy}, with different models considering the stellar mass ($M_\star$ ), the central velocity (VLSR), the flaring parameter ( $\psi$), the position angle (PA), and the source center ($x_0$, $y_0$) as fixed or free parameters.

Depending on the free parameters used (e.g., fixing the PA or allowing the fitting of vertical structure) and also on the velocity integrated image used, the stellar mass recovered would vary in the range of $0.16\sim0.24\,M_\odot$, which is consistent with a stellar mass derived from evolutionary models.\ However,  it is important to note that all the models and images would leave residuals, which span two times the channel velocity widths and are also strongly structured, as shown in Figure \ref{fig:MHO6_COmodel}.

To have a referential value for the stellar host mass, we used \texttt{CASA5.6} to get the position velocity diagrams (PV diagram from hereafter) along the major axis of each source, based on the position angle obtained from the continuum UV-modeling. The only exception is J0415, whose PV diagram was obtained along the east-west axis. The PV diagrams are shown in Figures \ref{fig:PV_subs} and \ref{fig:PV_nosubs} for the sources with continuum substructures and smooth profiles, respectively.

\section{Discussion}                    \label{sect:discussion}

\subsection{Detection of substructures}

We only detected obvious substructures in the brightest disks of our sample (CIDA\,1, MHO\,6, and J0433), which are also the most radially extended disks in gas and dust emission. The existence of strong dust traps located at larger radial distances from the star is most likely the reason for this observational result, as the dust is allowed to stay for longer timescales in the outer disk, thereby increasing the emitting area. On the other hand, our limited angular resolution only allowed us to detect substructures in the most extended disks.

This direct detection of substructures in 50\% of our sample does not represent the occurrence rate of substructures in all disks around VLMS, as our sample is biased towards the brightest disks and our spatial resolution is limited. Therefore, our observations only allowed us to directly detect deviations from a simple Gaussian profile in the disks where the extent of the emission is consistently larger than the synthesized beam size.

CIDA\,7 and J0420 are a good example of the limitations that our datasets have when detecting substructures. Even though  we are unable to confirm the existence of dust substructures in these systems, the  UV-plot in Fig.~\ref{fig:all_uvmodel} of CIDA\,7 shows some structure which is not described by a Gaussian profile, while  the residual image  of J0420 suggests that there might be more substructured dust emission that is not described by a single centrally peaked Gaussian in these disks . We lack the resolution and  sensitivity to confidently characterize them.
Given that these two disks are not totally dust depleted, the expected efficient radial drift must have been counteracted by a dust trapping mechanism. The compactness of these disks suggest that the dust trap is located so close to the star that our resolution did not allow us to detect it. 
In theory, any pressure bumps cannot be smaller than the local pressure scale height, which implies that if they are located closer to their star, their radial extent is smaller than our current resolution.
CIDA\,7  and J0420 are good candidates to be targeted by deep observations with ALMA in the most extended antenna configuration, allowing us to detect substructures of $\leq2\,$au in size at the distance of these targets, which is six and ten times smaller than the dust  $R_{68}$ of those disks, respectively. Future observations will test if even the very small disks around VLMS are able to generate dust traps, as it is observed in the massive and extended ones.

For J0415, our UV-coverage and sensitivity resolves the emission, and the centrally peaked Gaussian model does not leave any significant residual. Higher sensitivity is needed in order to discern deviations from the Gaussian profile (see Figure \ref{fig:all_uvmodel}).

\subsection{Origin of dust continuum rings and cavities}

All the detected substructures in our sample resemble ring-like emission. MHO\,6 is the only disk displaying what could be a hint of non-axisymmetric dust emission in the residuals at our current sensitivity. Rings structures are the most common type of substructure in moderate and massive stars, as shown by surveys such as DSHARP \citep{andrews2018, huang2018} and the Taurus survey at 1.3\, mm \citep{long2018}, and now a similar trend is found for the most extended disks around VLMS.

\subsubsection{Detected structures coincide with possible CO iceline location}

The iceline of each volatile marks the location in the disk where that volatile transitions from being mostly gas-phase to being frozen out on dust grains. It is possible that this phenomena could induce ring-like substructures in the dust continuum emission by changing the dust opacity and grain collisional fragmentation and growth rates \citep[e.g.,][]{okuzumi2016}.

To investigate if any of the iceline locations of the major volatiles coincide with the location of our gaps, we  calculated the midplane temperature of our disks by following \citet{kenyon1987}, where, for an irradiated flared disk, the temperature was parametrized as:
\begin{equation}
    T(r) = T_\star \, \left( \frac{R_\star}{r} \right)^{1/2} \, \phi_{\text{fl}}^{1/4} \text{,}
    \label{eq:temp_mid}
\end{equation}

\noindent with $r$ is the distance from the star,  $T_\star$ and $R_\star$ are the star temperature and radius, respectively, and $\phi_{\text{fl}}$ is the flaring angle, which we assume to be equal to 0.05 \citep{dullemond2004}. 
The stellar radius $R_\star$ was measured from the stellar luminosity $L_\star$ by assuming black body emission and spherical symmetry ($L_\star=4\pi R_\star^2 \sigma_{sb} T_\star^4$, with $\sigma_{sb}$ the Stefan-Boltzmann constant). As a first approximation, we ignored the contribution of accretion in our calculations of the stellar luminosity \citep[further discussed in][]{long2018}.
If the stellar radius is replaced in Equation (\ref{eq:temp_mid}), we can obtain the distance $r_i$ at which the temperature $T_i$ is reached, given a star of fixed luminosity $L_\star$, following:
\begin{equation}
    r_i = \frac{L_\star^{1/2}}{T_i^2}\, \cdot \, \left(\frac{ \phi_{\text{fl}} }{ 4 \pi \sigma_{sb} }\right)^{1/2} \text{.}
    \label{eq:iceline_rad}
\end{equation}

\begin{figure*}
 \centering
   \includegraphics[width=18cm]{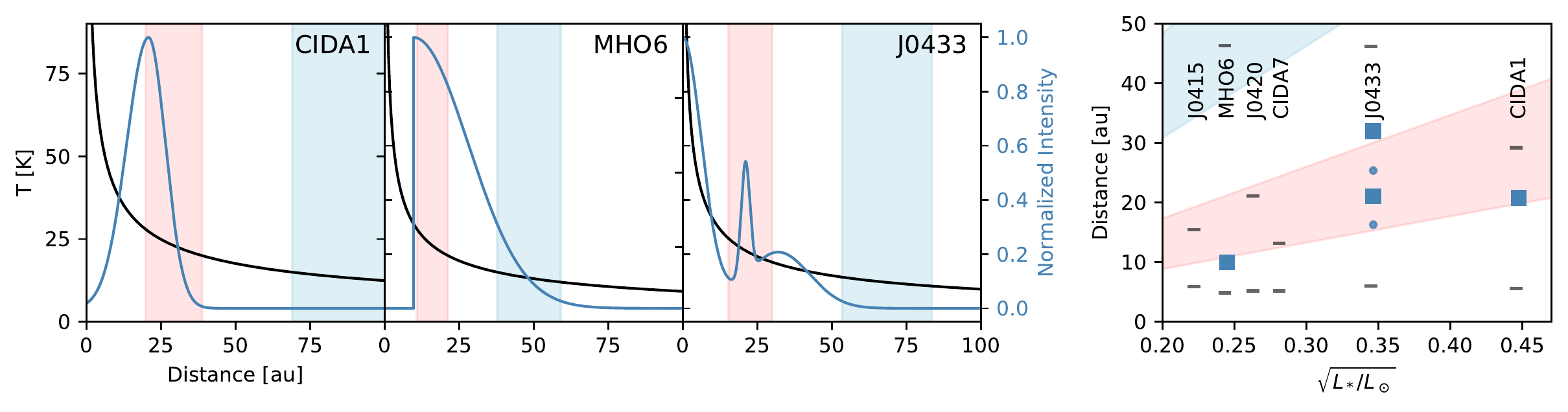}
   \caption{Left panels: Temperature profile of the midplane is obtained from equation (\ref{eq:temp_mid}) and plotted with a black curve. The blue curve shows the best radial profile obtained from the continuum UV-modeling, which was normalized to the peak emission. Vertical red and light blue shaded regions show the possible locations of the CO (light red) and N$_2$ (light blue) icelines. 
   Right panel: Disk radius is shown versus the square root of stellar luminosity ($\sqrt{L_\star}$). The shaded regions show the iceline location for CO (light red) and N$_2$ (light blue), obtained from equation (\ref{eq:iceline_rad}). The peak location of the modeled continuum emission rings are shown in squares, while the gap locations of J0433 are shown with dots. The lower line in each target marks half of the theoretical resolution given by the longest baseline for each dataset, as measured from the center, and the upper line marks the continuum $R_{90}$.}
   \label{fig:icelines}
\end{figure*}

We considered the two coldest icelines presented in \citet{zhang2015}, which are CO with a sublimation temperature of $\sim20\sim28$\,K (we took the lower limit $20$\,K from \citet{oberg2011} for direct comparison with \citet{long2018}), and the N$_2$ iceline which goes from $\sim12\sim15$K. The iceline location was calculated for each star using the parameters shown in Table \ref{tab:target_summary}, and the results are shown in Figure \ref{fig:icelines}.

In the right panel of Figure \ref{fig:icelines}, the positions of the detected substructures are displayed as a function of the root of stellar luminosity. Most of the substructures detected lie within the region where the CO iceline is identified. However, the role that the CO iceline plays in the morphology of our disk is not conclusive from our datasets. Even if we neglect the uncertainties introduced by the stellar parameters and the disk temperature, it is not clear if the structures that were generated by an iceline would result in a ring peaked at the iceline position or in a gap \citep{pinilla2017c, vandermarel2018}.

\citet{pinilla2017c} find that icelines do not strongly change the gas density around the icelines, nor can they carve a dust cavity as the ones we detected in MHO\,6 and CIDA\,1. Given that the peak of the rings detected in these disks are located very close to the position of the CO iceline, we cannot rule out the possibility that the iceline played a role in triggering the mechanism that is carving the cavity. In J0433, we find both peaks and gaps at the region (or close) where the CO iceline could be located. Given the radial compactness of the disks in our sample, the limited angular resolution, and the wide radial range covered by the CO iceline location, it is also possible that the substructures coincide with the iceline just by chance, in the same way that other surveys of substructures have not found a strong correlation between the position of substructures and iceline location \citep[e.g.,][]{long2018}. Additional deeper observations of the CO isotopologues could allow a better constrained modeling of the temperature of radial and vertical profile of these disks, thus providing better evidence for the role the iceline plays on the substructures detected.

\subsubsection{CIDA\,1}

Our modeling of the cavity in CIDA\,1 gives similar results to those found in \citet{pinilla2018b}, thus we do not further explore new possibilities for its cavity origin. It is important to note that our finding of the inner side of the ring being wider than the outer side is opposite to what we expected from the physical motivation of using this model, which accounts for the timescale of grain growth from micrometer to millimeter particles in dust traps \citep{pinilla2017a, pinilla2018b}. One possibility is that there is unresolved emission inside the main ring, and the inner side of the Gaussian is blending with it. Higher angular resolution observations are needed to understand the true nature of the dust distribution of this system.

\subsubsection{J0433}

If the gaps are assumed to be generated by a planet-disk interaction, we can use the width of the gap to estimate the mass of this gap-carving planet, under the assumption that this single planet is located at the position of the gap, and the ring is peaked at the local pressure maxima of the gas. In this scenario, if the physical parameters of the disk are kept constant, a more massive planet would create a wider gap \citep[e.g.,][]{fung2014, kanagawa2015, rosotti2016}.

For a crude estimate for the mass of the planet in the first J0433 gap, we followed a similar procedure as \citet{long2018}. In this approach, we assumed that the distance between the gap and ring scales with the Hill radius of the planet: 
\begin{equation}
R_{\text{Hill}} = r_p\, (M_p/(3\,M _\star))^{1/3} \text{,}
\end{equation} 
\noindent where $r_p$ is the location of the planet (the location of the gap coincides with the location of the planet, as the planet is carving the gap), $M_p$ is the mass of the planet, and $M_\star$ is the mass of the host star. If we consider the distance between the gap and ring to be $5\,R_{\text{Hill}}$  \citep[conservative upper limit for the gap width carved by a planet in][]{dodson2011}, and considering that our best model gives $r_{\text{ring}} - r_{\text{gap}} = 5$\,au for the first ring, then the approximate mass of the planet would be $M_p\sim0.1\,M_{\rm{Jup}}$. 

This calculation has a large uncertainty depending on the type of simulation. For instance, \cite{dodson2011} estimated a maximum of $4\,R_{\text{Hill}}$  between the planet location and the ring position, while \cite{pinilla2012b} estimated $7\,R_{\text{Hill}}$  for planets as massive as Jupiter. Taking these two limits ($4\,R_{\text{Hill}}$ and $7\,R_{\text{Hill}}$), the estimated planet mass becomes $M_p=0.11_{-0.07}^{+0.10}\,M_{\rm{Jup}}$. In this calculation, we did not account for the physical conditions of the disk, such as turbulence or temperature, nor did we consider the minimum mass to open a gap in the disk.\ Therefore, it just gives us a crude estimate for the mass order of magnitude of a single planet carving the gap. A future analysis should consider higher angular resolution observations to better constrain the gap-ring morphology as well as dedicated hydro-simulations to estimate the planet candidate mass.

\subsubsection{MHO\,6}

In MHO\,6, the modeling indicates the existence of a central cavity that is slightly asymmetric. Several processes in protoplanetary disks could create this type of an inner cavity, such as photoevaporation due to stellar irradiation \citep[e.g.,][]{alexander2007, owen2012, owen2019}, companions \citep[e.g.,][]{price2018},  a planet-disk interaction \citep[e.g.,][]{rice2006, zhu2011}, and dead zones \citep[e.g.,][]{flock2015}. In the following, we discuss each one of these scenarios for the MHO\,6 cavity.

\paragraph{Photoevaporation:} The $10\,$au dust cavity and the low accretion rate \citep[{}$5\times10^{-11}\,\text{M}_\odot\,\text{yr}^{-1}$ estimated from
a UV excess in][]{herczeg2008} are in agreement with the predicted evolution for a very low mass star ($0.1\,\text{M}_\odot$) having its material stripped away by a photoevaporative flow \citep{owen2012}. However, we do not see clear evidence of a gas depleted cavity in the $^{12}$CO emission, although the $^{13}$CO emission suggests a decrease in the gas density in the inner region. A follow-up with dedicated VLMS photoevaporation models and higher angular resolution of this target in different molecular lines would be able to characterize the impact of this mechanism in the cavity we detect. From the current observations, this mechanism is still an option for the observed cavity.

\paragraph{Companion:} Binary companions produce cavities in circumbinary disks, with sizes in the range of $3 - 5$ times the binary semi-major axis, depending on the mass ratio, the eccentricity, and the disk viscosity \citep{artymowicz1994, miranda2017b, ragusa2017}. They also produce quasi-periodic variations in the accretion rate \citep{munoz2016}. If MHO\,6 was a circumbinary disk, the location of this companion could not be farther out than $\sim3\,$au ($\sim20\,$mas) from the primary star. However, previous independent observations of MHO\,6 have not found any evidence of multiplicity in this system \citep[][the latest reference identified companions in its survey as late-type as M0.0]{briceno1998, kraus2007, herczeg2014}. 
A spectroscopic follow-up with radial velocities, given the high inclination of the system, could be useful to constrain the upper mass limit of a companion in the close inner region.

\paragraph{Embedded planets:} 
The planet population around very low mass stars spans a wide range of masses, where even giant planets have been detected \citep[e.g.,][{}a 0.46 $M_{\rm{Jup}}$ around a 0.1\,$M_\odot$ VLMS]{morales2019}, meaning that protoplanetary disks around VLMS probably have the potential to create such objects. If we make a simple calculation following the opening gap criterium introduced by \cite{crida2006}, the approximate mass needed for a single planet to open a gap and explain the cavity observed in MHO\,6 is 1 or $0.6\,M_\text{Saturn}$ ($0.18$ and $0.3\,M_{\rm{Jup}}$), located at $\sim$7\,au for $\alpha=10^{-3}$ and $\alpha=10^{-4}$, respectively. If we assume a gas to dust ratio of 100,
we estimate a disk mass of $\sim12.6\,M_\text{Saturn}$ or $\approx 3.8\,M_{Jup}$ given the dust mass
obtained in Section \ref{sec:dust_mass}.\ This means that the cavity opening planet is $<10\%$ of the current estimated mass of the disk. This implies that this type of potential single planet could have formed within the disk of MHO\,6, although we do not exclude the possibility of multiple planets being responsible of this central cavity. 

Another characteristic to take into consideration is the location of the peak brightness in the azimuthally averaged radial profile of $^{13}$CO and dust continuum emission. 
We find hints of the cavity in the $^{13}$CO line emission with its peak at 10\,au when measured from the image. The same peak appears at $14\,$au when measured from the dust continuum image. Although these comparisons are strongly biased by the beam shape, it supports the idea that a planet may be responsible for the cavity formation since models predict such segregation between the outer edge of the gap in gas versus dust \citep{dejuanovelar2013, facchini2018}. Future higher angular resolution observations could test this hypothesis by better characterizing the CO brightness profile.

\paragraph{Dead zone:} A dead zone is a low-ionized region at the disk midplane, where the dense environment blocks the high energy radiation, suppressing the magneto-rotation instability and, therefore, the angular momentum transport. It has been shown that the presence of dead zones can open gaps and cavities by forming gas pressure bumps at the outer edge of the dead zones, where dust trapping is efficient \citep[e.g.,][]{regaly2012, flock2015}. \citet{pinilla2016} predict that cavities that formed by dead zones alone would have millimeter- and micrometer-sized particles concentrated at the peak of the gas density. As a result, the radial location of the ring in millimeter wavelengths and scatter light would be the same. If we neglect temperature effects, our images suggest that the peak of $^{13}$CO is closer to the star than the millimeter peak. 
Therefore, we expect for the micrometer-sized particles to peak closer to the star as well. 
In the dead zone scenario, \cite{pinilla2016} also predict a strong gas depletion in the outer parts of the disk, further out from the dust trap, which is not currently seen in our observations of $^{13}$CO. 
If a magnetohydrodynamical wind is included in the models together with a dead zone, the observational diagnostics are very similar to planets. One step forward to try to disentangle between these models is to image this disk in scattered light and search for potential planets in the cavity. However, these disks are too faint to image given the limitations of current telescopes in the optical and near-infrared.

\bigskip

To summarize, the formation of the cavity observed in MHO\,6 could be explained by one or a combination of the mechanisms we have discussed above. Several observational efforts can be done to disentangle these possibilities, such as a better estimate of the star accretion rate, a deeper search for planets or companions in its cavity in the optical and infrared wavelengths as well as imaging the disk in this regime, and very deep and higher angular resolution observations of molecular lines.

\subsection{MHO\,6 kinematics} \label{sec:mho6kinematics}

Given that MHO\,6 is the only disk with the gas emission detected across the whole velocity range, it was a good candidate for the dynamical mass measurement of its star. After several attempts to model MHO\,6 rotation with different images, masks, and a combination of free parameters, we find our range of recovered stellar masses (between $\sim0.16-0.24\,M_\odot$ ) to be consistent with the values from pre-main sequence models of MHO\,6 \citep[$0.09 - 0.20M_\odot$][]{kraus2009, herczeg2014, wardduong2018}, but all the kinematic models would leave strong structured residuals with an amplitude spanning two times the channel velocity width.

Although our observations have enough spatial resolution to resolve the vertical structure in some of our disks (as seen in Figure \ref{fig:gas_subs} as a cone-like emission in the $^{12}$CO moment 0 of MHO\,6 and J0433), the low S/N did not allow us to differentiate between the emission coming from the back side of the disk from that of the front side. The integrated velocity map contains the emission of both sides as if they were the same, and thus parameters such as $M_\star$ or $\psi$ from $Z(r, \psi) \propto r^{\psi}$ could not be recovered reliably.

This issue is not an exclusive problem in VLMS disks, but it applies to all the gas measurements with a high angular resolution and poor S/N. Future approaches to accurately recover $M_\star$ should consider more robust methods, such as UV-modeling \citep[similar to the \texttt{DiskJockey} code from][]{czekala2015}.

MHO\,6 is a good candidate to further study the kinematics, substructures, and the physics of planet formation in VLMS. Its brightness allows high S/N observations at high angular resolution with non-prohibitive integration times. Combining the datasets presented in this paper, plus an observation of $\geq 5\,$hr of time on source with ALMA using long baseline configurations should have enough resolution and sensitivity to precisely characterize the kinematics of the CO isotopologues, as previously done with disks around T-Tauri and Herbig stars \citep[e.g.,][]{pinte2019, teague2019}.

\subsection{Comparison of dust  $R_{68}$ and $L_{\rm{mm}}$ with previous studies}\label{sec:differences}

All our sources have observed or estimated 0.87\, mm fluxes, and a subset also had their dust continuum  $R_{68}$ measured with previous UV-modeling. In the following, we compare and discuss our measurements with previous results, and we include the VLMS in the study of a size-luminosity relation.

\paragraph{Flux:} J0415 was observed using the Submillimeter Array (SMA) at a wavelength of 1.3\, mm, with a reported flux density of $12.6\pm1.4\,$mJy in \citet{andrews2013}, where they extrapolated this measurement to $0.87\,$ mm using $F_\nu \propto \nu^\alpha$, with $\alpha=2.4\pm0.5$, thus obtaining $F_{0.87\rm{mm}}=32.9\pm15.2\,$ mJy. Although for all our other five sources, the extrapolations from the SMA flux measurements are consistent within the error range, the flux at 0.87\, mm received from J0415 is approximately 35 times dimmer than expected, as observed independently by two different ALMA projects. 
A possible explanation could be a sudden change in millimeter flux due to flares, as it has been observed in low mass stars \citep[e.g.,][]{macgregor2018}. 
For now, the only ALMA observations of this disk are the ones presented in this work, which are both in Band 7. Future observations at 1.3\, mm should solve this discrepancy with the SMA results.

\paragraph{Dust $R_{68}$:} The radius enclosing 68\% of the dust continuum  emission ($R_{68}$) was previously estimated for MHO\,6 and J0433, using UV-modeling on SMA 340\,GHz observations. In \citet{tripathi2017} and \citet{andrews2018a}, they estimated a value of $R_{68}=36.9_{-5.7}^{+8.5}\,$au for MHO\,6 (${0.26''}^{+0.06}_{-0.04}$), and $R_{68}=58.9_{-8.7}^{+6.9}\,$au for J0433 (${0.34''}^{+0.04}_{-0.05}$), which are consistent with the values measured by this work. Both are slightly overestimated (1.6 times for J0433) compared to the values of this work due to the considerably lower angular resolution of SMA observations compared to ALMA ($0.86''\times0.80''$ for MHO\,6 and $0.61''\times0.52''$ for J0433).

\paragraph{$R_{68} - L_{\rm{mm}}$ relation:}

\begin{figure}
 \centering
   \includegraphics[width=9cm]{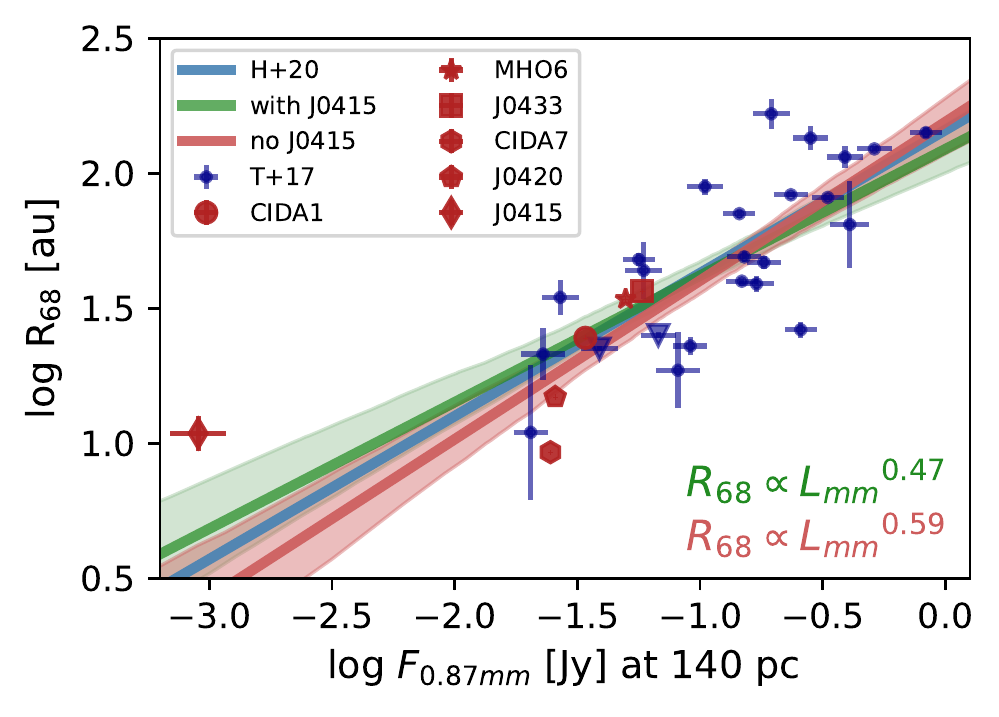}
   \caption{Relation between $\log(R_{68})$ and $\log(F_{0.87\rm{mm}})$. The points in blue are from \citet{tripathi2017}, with distances scaled to $140\,$pc using distances inferred from Gaia DR2, similar to \citet{andrews2018a}. The solid line in blue is the linear regression found by \citet{hendler2020} (abbreviated as H+20), fitting the data from \citet{tripathi2017}. The VLMS from this work are shown in red. The bests linear regression fits and their $68\%$ confidence intervals are plotted in red when J0415 is excluded from the fit, and in green when all points are considered.
   \label{fig:lmm_reff}}
\end{figure}

We combined our VLMS measurements with the Taurus sample observed with the SMA at 340GHz from \citet{tripathi2017} in order to compare this all with the recent analysis by \citet{hendler2020}, where they obtained a relation of $R_{68}\propto {L_{\rm{mm}}}^{0.53}$ for the Taurus star-forming region. For comparison purposes with this study, we followed the same approach using the Bayesian linear regression described in \citet{kelly2007}, which was implemented in the python package \texttt{linmix} \citep[publicly available in github, see][]{linmix2015}, to fit a linear relation between $R_{68}$ and $L_{\rm{mm}}$ following:
\begin{equation}
    \log_{10}(R_{68}) \, = \, \alpha \, + \, \beta \, \log_{10}(L_{\rm{mm}}) \text{,}  \label{eq:linear_rel}
\end{equation}

\noindent where $\alpha$ and $\beta$ are the regression coefficients. Our best fit was calculated by using the median value of the last 200000 steps after convergence. We find that including our VLMS sample does not statistically change the previous result. However, the inclusion of J0415 changes the steepness of this relation in about $1\,\sigma$ . 
If we consider J0415 as part of the fitting data, then the relation recovered is $\alpha=2.09\pm 0.09$, $\beta=0.47\pm0.08$. 
If we exclude J0415 from the fitting, the result is $\alpha=2.19\pm 0.10$, $\beta=0.59\pm0.10$. 
Both relations are close to the $1\,\sigma$  limit of each other, and they also overlap with the previous calculation from \citet{hendler2020}, as shown in Figure \ref{fig:lmm_reff}. 
If we check their relations' intrinsic scatter from the linear regression, we obtain $\sigma_{\text{scatter}}=0.231\pm0.041$ and $\sigma_{\text{scatter}}=0.217\pm0.040,$ respectively. Therefore, including J0415 does not significantly increase the scatter.

The difference in the steepness of the recovered relations is not statistically significant from what was previously found. 
However, it is not completely clear if the same single power law relation between the millimeter luminosity and the size of the disks holds along the whole luminosity range. 
Given that J0415 is the only source with its size measured in the $\sim1\,$ mJy brightness range, it is unknown if disks have some mechanisms to remain extended even when they are low in dust content (thus flattening the relation between size and luminosity in the low luminosity regime), or if a J0415-extended dust size is part of the relation scatter that is also observed in bright disks. 
To understand if J0415 is an outlier and to test if the power law behavior of the $R_{68} - L_{\rm{mm}}$ relation flattens or holds at the low brightness regime, we need more deep observations at a high angular resolution of disks with $F_{0.87\text{mm}}<10\,$ mJy. This could be achieved by observing each source from several tens of minutes to a few hours in ALMA Band 7.

\subsection{J0415 dust radial extent}

Although CIDA\,7 has a dust content of at least $\sim 27$ times higher than J0415, it remains an open question as to how both can have a similar size (see Table \ref{tab:rgas_rdust}). 
This result could be due to weaker dust traps in J0415 compared to those in CIDA\,7, thus CIDA\,7 can trap the dust more efficiently.

Alternatively, if the $M_{\rm{gas}}$ of the disk is very low, the millimeter grains would have a high Stokes number and radial drift would become negligible. In \citet{pinilla2017b}, they explore this scenario with a disk of $60$\,au in radius around a BD of $0.05M_\odot$. When $M_{\rm{gas}} = 2\cdot10^{-2}\,M_{\rm{Jup}}$, diffusion and drift still depleted the disk from millimeter particles; however, when $M_{\rm{gas}} = 2\cdot10^{-3}\,M_{\rm{Jup}}$, the millimeter grains were decoupled from the gas, and they could remain in the disk for longer timescales. For J0415, if we assume a dust to gas ratio of 1/100, we obtain $M_{\rm{gas}} \approx4\cdot10^{-2}\,M_{\rm{Jup}}$, which does not seem to be low enough for a complete dust-gas decouple to occur. Additional observations are needed in order to characterize its gas radial density profile and to test this possibility.

Finally, it has been proposed that dust grains could be growing in a fractal manner, such that large aggregates would avoid radial drift by maintaining a low Stokes number \citep{kataoka2013}. This scenario can be distinguished from the compact millimeter grains by measuring the opacity index $\beta$ between the $1$\,mm and $3\,$mm emission \citep{kataoka2014}. Those observations, however, would require a very high sensitivity with enough angular resolution to spatially resolve the radial profile of J0415.

\subsection{$R_{\rm{gas}}/R_{\rm{dust}}$ ratio} \label{sec:rgas-rdust}

\begin{figure}
 \centering
   \includegraphics[width=9cm]{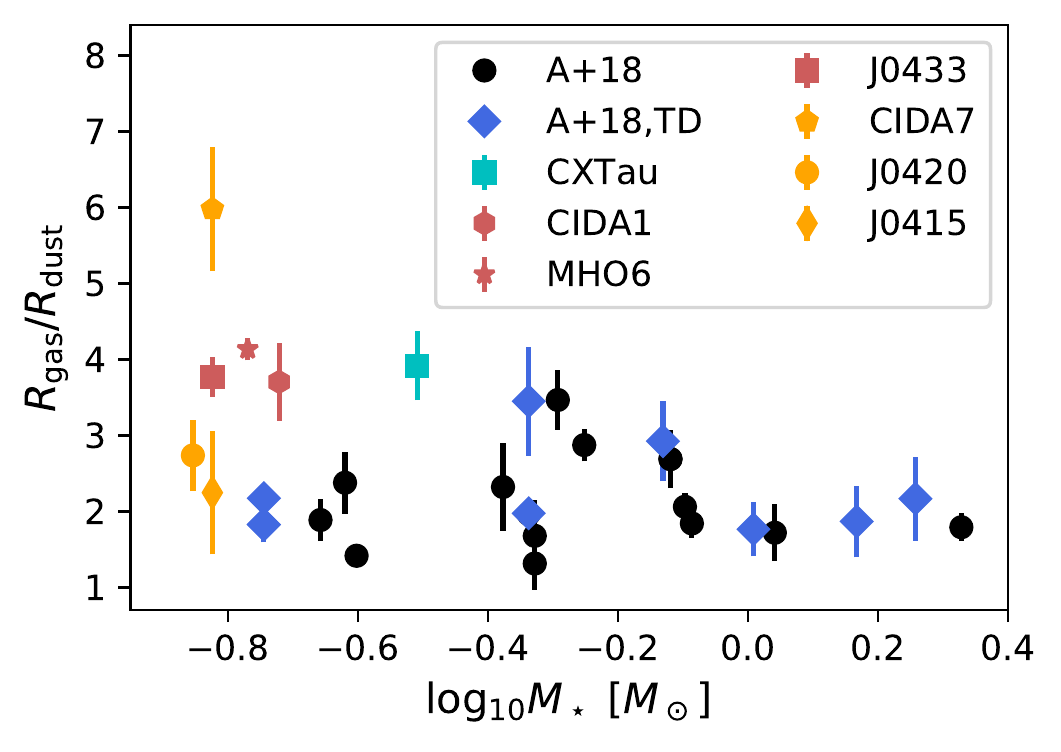}
   \includegraphics[width=9cm]{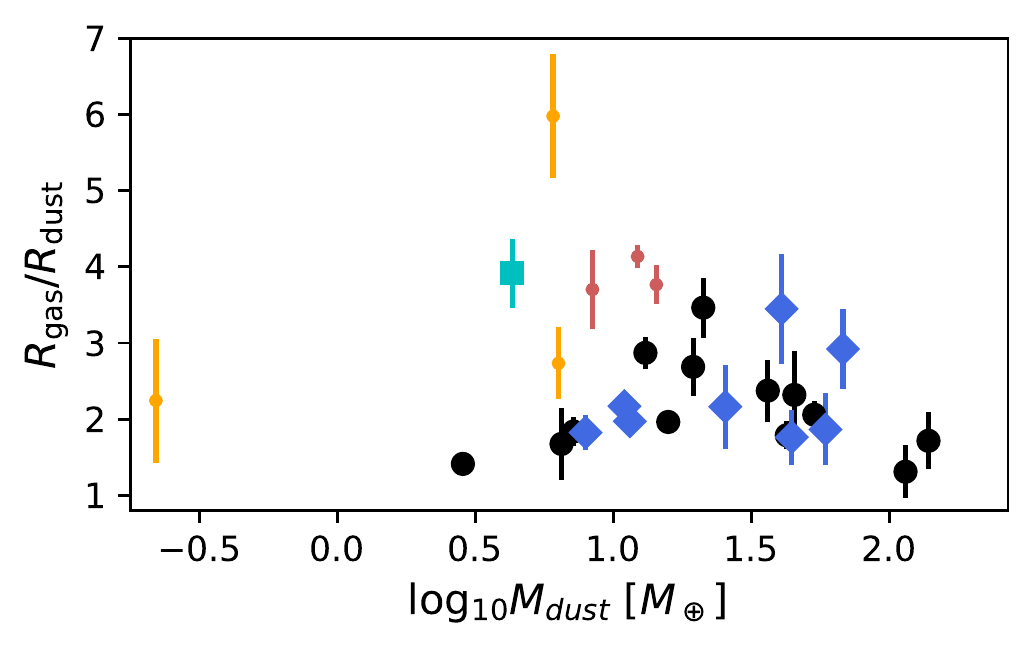}
   \caption{Upper panel:  $R_{\rm{gas}}/R_{\rm{dust}}$ as a function of the stellar mass of the disks in the Lupus star-forming region (SFR)  reported in \cite{ansdell2018} (abbreviated as A+18), CX\,Tau \citep{facchini2019}, and the targets reported in this paper. For CX\,Tau, we took the $R_{90}$ radii of the dust and gas. The VLMS of this work are shown in red for the sources with a substructure, and they are in yellow for the smooth sources, with the values from Table \ref{tab:rgas_rdust}. Ratios from the Lupus SFR by \citet{ansdell2018} were calculated from $R_{90}$ and a different color was used for disks identified as transition disks (TD). Lower panel: $R_{\rm{gas}}/R_{\rm{dust}}$ of the same targets, but compared to the dust mass of the disks. The dust mass was calculated from Equation (\ref{eq:mdust}) under the assumption that $T=20$\,K for the whole disk midplane, with distances from GAIA DR2. }
   \label{fig:rgas_rdust}
\end{figure}

We find that four of our six disks show a ratio between the $R_{90,\rm{gas}}$ and $R_{90,\rm{dust}}$ that is very close to or above $4.0$, which is similar to the ratio measured in CX\,Tau \citep{facchini2019} (ratio of $3.9\pm0.5$ at $R_{90}$), as compared in Figure \ref{fig:rgas_rdust}. 
Our values, however, are conservative measurements of radii ratios for the sources with strong cloud contamination (all the sources in our sample, except for MHO\,6), and so they should be considered as a lower limit. 
As it can be seen in the velocity channel maps in Figures \ref{fig:chanmap_cida1}, \ref{fig:chanmap_j0433}, \ref{fig:chanmap_cida7}, \ref{fig:chanmap_j0420}, and \ref{fig:chanmap_j0415}, the channels that are less affected by cloud contamination are the highest velocity channels, which are closer to the star. 
As soon as the emission gets radially extended in the channels closer to the central velocity, the cloud extinction becomes so high that we cannot see the disk emission anymore. 
Our integrated flux moment 0 images and the gas emission are then biased towards the compact emission located close to the star, underestimating the $R_{\rm{gas}}$ measurement. This effect is particularly strong in J0420 and J0415, as they are the dimmest sources in this work, and J0420 also has the highest extinction (see Fig.~\ref{fig:contamination}).

Our observations show that it is common for bright disks around VLMS to have a gas radial extension of $>3$ times the millimeter dust radial extension, which is consistent with the efficient radial drift expected for millimeter-sized grains in VLMS disks \citep{pinilla2013, zhu2018}.
In fact, recent thermochemical modeling including dust evolution by \citet{trapman2019, trapman2020} shows that ratios of $R_{\rm{gas}}/R_{\rm{dust}}>4$ cannot be solely explained by effects of optical depths; additionally, in those cases, radial drift is required to explain the gas and dust size difference. 
Despite the limited sensitivity and cloud contamination, our disks are still very close and even above that limit. 
Although similar ratios have been observed in sources with a moderate stellar mass \citep[$\sim0.5\,M_\odot$, see ][]{ansdell2018}, as shown in the upper panel of  Figure \ref{fig:rgas_rdust}, the direct comparison between our works is hindered by the data differences, such as in the sensitivity and angular resolution, and also by the different approaches used to obtain the gas and dust radii: Our continuum radial extension was calculated from the visibilities rather than the images, and we did not apply Keplerian masking in the flux integrated gas images. 
A more complete sample, which includes more observations of disks around very low and moderate mass stars, and also uniform data sensitivity and analysis are required to confirm if there is a general trend where $R_{\rm{gas}}/R_{\rm{dust}}$ increases towards lower stellar masses.

According to \citet{trapman2019} and \citet{trapman2020}, they expect more massive disks to have a higher $R_{\rm{gas}}/R_{\rm{dust}}$ ratio, driven by a larger observed $R_{\rm{gas}}$ due to the greater total CO content, and also because the higher dust content would produce a more efficient grain growth and inward radial drift. However, we do not observe this trend in the lower panel of Figure \ref{fig:rgas_rdust}. Apparently, a decreasing radii ratio is obtained towards higher dust mass disks, which could be the result of an efficient radial drift in disks around VLMS, and the linear relation between $\log(M_\star)-\log{M_{\rm{dust}}}$. This supports the idea that smaller disks result from a fast radial drift \citep{long2019} due to their inability to trap dust in the outer regions, while more massive disks are more capable of creating dust traps father away from the central star.

CIDA\,7  stands out in our sample as having the most extreme $R_{\rm{gas}}/R_{\rm{dust}}$ ratio observed so far, with the gas being six times more extended than the dust, which is well beyond the ratio of four limit from \citet{trapman2019}. 
This confirms that radial drift is responsible for the compact size of this source. However, it is not completely dust depleted, so radial drift has been counteracted by another mechanism, or a combination of them. 
The southern non-Keplerian emission detected in this system (in the velocity channels $4.4$ and $4.8\,$km\, s$^{-1}$ from Fig.~\ref{fig:chanmap_cida7} and \ref{fig:gas_subs}) was masked when measuring the $R_{90, \rm{gas}}$, 
so the ratio of six is between the disk rotating gas and the dust size recovered from the model. We were unable to determine the origin of this extended emission in the south, as it is only detected in two different channels and it does not appears to be axisymmetric. 
A multiwavelength follow-up, with high sensitivity and angular resolution, might be required to understand the nature of this emission, as it could be explained by several different mechanisms, such as winds, outflows, interactions with external companions, an interaction with the surrounding cloud and envelope, among others.

The lowest gas-to-dust size ratio in our sample, measured in J0415, is likely due to the combined effects of lower than expected brightness, a low sensitivity due to our inability to apply self-calibration, and the extinction due to cloud contamination. Although we were unable to confidently recover the gas radius, our observations set a lower limit for its radial extension, and we confirm that in this very low disk mass regime, the gas emission is still more extended than dust emission. A more precise measurement of the gas radius requires a combination of deeper observations of lines less affected by the surrounding cloud and envelope as well as line modeling.

\section{Conclusions}                   \label{sect:conclusions}

To understand the process of planet formation in VLMS and compare it with our current knowledge of planet formation in solar type stars, we observed and studied a sample of the brightest disks around VLMS in Taurus, at a $0.1''$ resolution and at a $0.87\,$mm wavelength. This sample is composed of CIDA\,1, MHO\,6, J0433, CIDA\,7, J0420, and J0415 (2MASS  names in Section \ref{sect:obs}). Here we summarize our main conclusions as follows.

\begin{itemize}
    \item  \textbf{Detection rate of substructures:} Millimeter dust substructures were directly detected in only 50\% of the targets in our sample. Our results suggest that the detection of substructures in disks around VLMS is limited by angular resolution and sensitivity, since the dust radial extent is very small and these disks are also very faint. Deep, high angular resolution observations over a non-brightness biased sample of VLMS should confirm the ubiquity of substructures in these disks. 

    \item \textbf{Substructured disks:} Substructures were detected in CIDA\,1, MHO\,6, and J0433; with the latest two being new detections. These three disks are the brightest and largest in our sample. They all  have axisymmetric ring-like substructures, and only MHO\,6 shows a weak asymmetry of amplitude less than 5\% of the peak brightness. 
    Both CIDA\,1  and MHO\,6 show central cavities in their emission. 
    If we assume that a planet-disk interaction is the origin of the MHO\,6 cavity, then a Saturn-mass planet (0.3\,$M_{\rm{Jup}}$) is needed \citep[as in the case of CIDA\,1][]{pinilla2018b}. This planet should be located around 7\,au. However, we cannot exclude other mechanisms that can also explain the origin of this cavity, such as multiple planets,  a dead zone, a binary companion, or photoevaporation. 
    Our UV-modeling of J0433 suggests that this disk could have two rings located at 21 and 32 \,au. However, we cannot confirm the separation between both at the $1\,\sigma$  errorbar.
    We estimate that a planet of $\sim0.1\,M_{\rm{Jup}}$ in mass could explain the first gap-ring.
    The substructures were detected within the region where the CO iceline could be located. Our current datasets lack the necessary S/N and resolution to properly characterize the vertical and radial temperature profile of the CO isotopologues, and so future deeper observations will be needed to determine if the iceline played any role in triggering or maintaining the substructures observed.\\

    \item \textbf{Smooth disks:} The dust disks in CIDA\,7, J0420, and J0415 are the less radially extended, less massive disks of our sample. With an angular resolution of $0.1''$, these disks  are well described by a single Gaussian radial profile, which we used to measure their sizes.
    CIDA\,7  is the most compact of them, with an $R_{90}=13.16\,$au, which is similar to the $15.46\,$au from J0415. Yet, the dust mass estimate suggests that CIDA\,7  is $\sim27$ times more massive. 
    In J0420, the residual continuum image shows some structured non-axisymmetric emission with $5\,\sigma$  peaks. However, this emission is very low in contrast to the smooth emission, which is over $300\,\sigma$  at its peak. Higher  angular resolution observations are needed to describe the potential substructures in these disks .\\

    \item \textbf{Size-luminosity relation:} The disks in our sample follow a similar relation between $L_{\rm{mm}}-R_{68}$ as the one found for bright disks in the same star-forming region \citep[see][]{hendler2020}. However, our single measurement of a disk size in the low luminosity regime (J0415) needs to be complemented with deeper additional observations of other sources with a low stellar mass and low disk brightness. 
    These measurements will help us understand the behavior of the size-luminosity relation across the whole range of disk sizes, enabling us to test if a single power law describes it. \\

    \item \textbf{Evidence of efficient radial drift:} When considering the dust and gas radii as the location where 90\% of the emission is enclosed,  four out of six disks in our sample show a ratio between $R_{\rm{gas}}/R_{\rm{dust}}$ above 3.5. 
    This is expected for disks where radial drift is depleting the dust, suggesting that radial drift is more efficient in VLMS than in moderate or high mass stars. 
    However, the analysis of the sizes of disks in Lupus by \citet{ansdell2018} also allowed them to find moderate stellar mass sources with similar disk size ratios as those observed in VLMS, which suggest that we need more observations to confirm this trend. 
    Our comparison with \citet{ansdell2018} does not account for the differences in data acquisition and $R_{90}$ calculation, which should be considered in future works, aiming for a uniform analysis of the extension of disks between different star-forming regions. The most extreme case of high $R_{\rm{gas}}/R_{\rm{dust}}$ in our VLMS is observed in CIDA\,7, with a value of six. This very high $R_{\rm{gas}}/R_{\rm{dust}}$ ratio suggests that strong radial drift is at play \citep{trapman2019}, raising the question about how this disk remains massive in dust.

\end{itemize}

Our observations do not exclude giant planet formation as an explanation for the substructures detected.  Disks around VLMS follow similar trends as those that have been observed in disks around higher mass stars, based on our sample of bright disks. A future confirmation of a deviation from current correlations of physical parameters will require the recalculation of fluxes and sizes of the Taurus disks by using more sensitive and higher angular resolution observations from ALMA in a larger and more complete sample.

\section*{Acknowledgements}
We thank Timmy Delage and Reema Joshi for their very helpful comments. We also thank Sean M. Andrews for the insightful discussions, which significantly improved this manuscript.
N.K. and P.P. acknowledge support provided by the Alexander von Humboldt Foundation in the framework of the Sofja Kovalevskaja Award endowed by the Federal Ministry of Education and Research.
This Project has received funding from the European Union's Horizon 2020 research and innovation programme under the Marie Sklodowska-Curie grant agreement No 823823 (DUSTBUSTERS). This work was partly supported by the Deutsche Forschungs-Gemeinschaft (DFG, German Research Foundation) - Ref no. FOR 2634/1 TE 1024/1-1.
This paper makes use of the following ALMA data: ADS/JAO.ALMA\#2012.1.00743.S, ADS/JAO.ALMA\#2015.1.00934.S, ADS/JAO.ALMA\#2016.1.01511.S, and ADS/JAO.ALMA\#2018.1.00310.S.  ALMA is a partnership of ESO (representing its member states), NSF (USA) and NINS (Japan), together with NRC (Canada), MOST and ASIAA (Taiwan), and KASI (Republic of Korea), in cooperation with the Republic of Chile. The Joint ALMA Observatory is operated by ESO, AUI/NRAO and NAOJ.

\bibliographystyle{aa}

\begin{appendix} 
\section{Summary of ALMA observations}

\begin{figure}
 \centering
        \includegraphics[width=9cm]{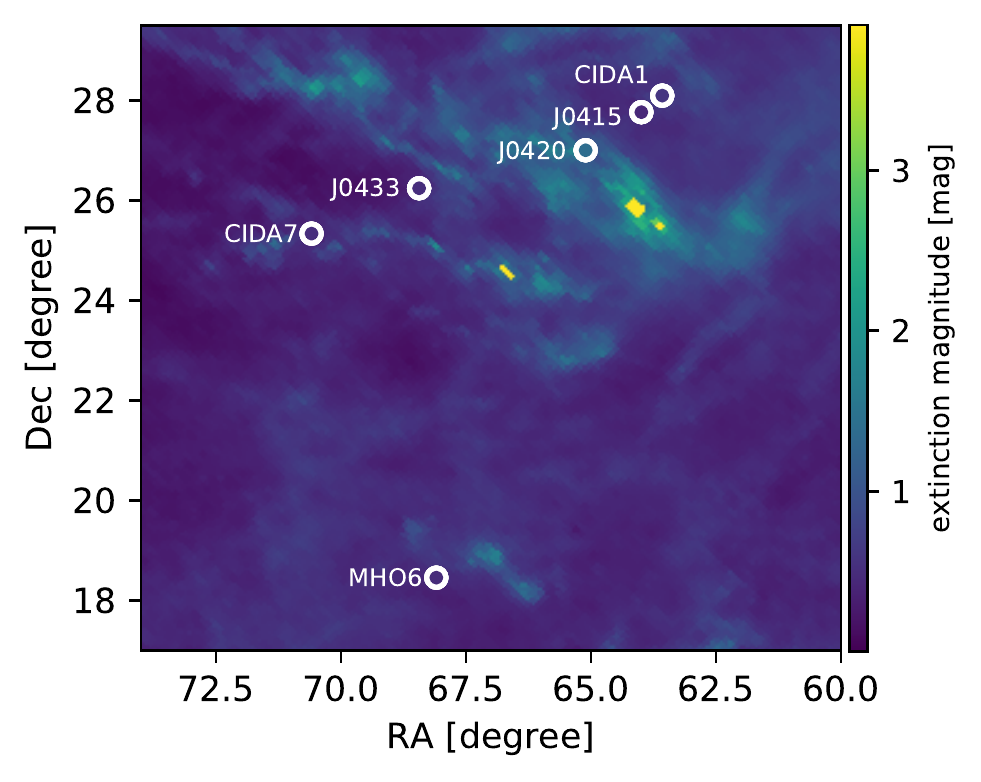}
   \caption{Spatial distribution of the Taurus sample studied in this work. The background is an extinction map compiled by \citet{schlafly2014}.}
   \label{fig:contamination}
\end{figure}

A summary of the observation log is provided in Table \ref{tab:obs_log}. The difference for configurations ``Compact'' and ``Extended'' denotes the stage at which we started its self-calibration based on the spatial extension of the antenna array, see Section \ref{sect:obs_reduction}. The properties of the dust continuum emission images and gas emission datacube are shown in Table \ref{tab:imaging_cont} and \ref{tab:imaging_CO}, respectively. In Figure \ref{fig:contamination} we show the reddening map from \citet{schlafly2014} of the Taurus star-forming region, with the location of our sources. The reddening values for the sources are 0.61 for CIDA\,1, 0.48 for MHO\,6, 0.51 for J0433, 0.50 for CIDA\,7, 1.38 for J0420, and 0.44 for J0415.

\begin{table*}
\begin{center}
\caption{Summary of ALMA observations. \label{tab:obs_log}}
\begin{tabular}{ |l|c|c|c|c|c|c| } 
    \hline
    \hline
    Source  & Program ID     &  Obs. Date  & Exp. time & N$^\circ$ & Baselines & Configuration \\
            &                &             & (min)     & Antennas  & (m)       &               \\
    \hline
    CIDA\,1    & 2015.1.00934.S &  2016-08-12 & 47.68     & 38        & 15 - 1462 & Compact   \\
            & 2016.1.01511.S &  2017-07-06 &  4.23     & 42        & 17 - 2647 & Compact   \\
    \hline
    MHO\,6    & 2012.1.00743.S &  2013-11-19 &  4.66     & 28        & 17 - 1284 & Compact   \\
            &                &  2014-07-27 &  2.71     & 33        & 24 - 820  & Compact   \\
            & 2018.1.00310.S &  2019-08-21 & 47.15     & 45        & 41 - 3189 & Extended  \\
    \hline
    J0433   & 2012.1.00743.S &  2013-11-17 &  2.80     & 29        & 17 - 1284 & Compact   \\
            &                &  2014-07-27 &  2.27     & 33        & 24 - 820  & Compact   \\
            & 2018.1.00310.S &  2018-11-20 & 12.98     & 47        & 15 - 1398 & Compact   \\
            &                &  2018-11-24 & 12.98     & 46        & 15 - 1261 & Compact   \\
            &                &  2019-08-13 & 26.61     & 43        & 41 - 3144 & Extended  \\
            &                &  2019-08-13 & 10.45     & 43        & 41 - 3144 & Extended  \\
    \hline
    CIDA\,7    & 2018.1.00310.S &  2018-11-13 & 12.48     & 47        & 15 - 1398 & Compact   \\
            &                &  2019-08-24 & 26.04     & 47        & 41 - 3396 & Compact   \\
            &                &  2019-08-24 & 26.04     & 47        & 41 - 3638 & Extended  \\
    \hline
    J0420   & 2012.1.00743.S &  2013-11-19 &  4.66     & 26        & 17 - 1284 & Compact   \\
            &                &  2014-07-27 &  2.27     & 33        & 24 - 820  & Compact   \\
            & 2018.1.00310.S &  2018-10-30 & 12.98     & 48        & 15 - 1398 & Compact   \\
            &                &  2018-11-13 & 12.98     & 47        & 15 - 1398 & Compact   \\
            &                &  2019-08-12 & 27.22     & 47        & 41 - 3638 & Extended  \\
            &                &  2019-08-13 & 27.22     & 43        & 41 - 3144 & Extended  \\
            &                &  2019-09-18 & 27.25     & 41        & 15 - 3638 & Extended  \\
    \hline
    J0415   & 2016.1.01511.S &  2015-09-20 & 4.8       & 42        & 15 - 3189 & Extended  \\
            & 2018.1.00310.S &  2019-09-20 & 27.88     & 45        & 15 - 3189 & Extended  \\
    \hline
    \hline
\end{tabular}
\end{center}
\end{table*}

\begin{table*}[ht!]
\begin{center}
\caption{Continuum imaging summary. \label{tab:imaging_cont}}
\begin{tabular}{ |c|c|c|c|c|c|c|c| } 
    \hline
    \hline
    Source  & RA           & Dec             & Beam                   & Peak $I_\nu$       & RMS noise          & $F_{\rm{0.87mm}}$ & $M_{\rm{dust}}$ \\
            & (ICRS)       & (ICRS)          & (mas$\times$mas, deg)  & (mJy\,beam$^{-1}$) & (mJy\,beam$^{-1}$) & (mJy)             & ($M_\oplus$) \\
    \hline
    CIDA\,1    & 04:14:17.620 & $+$28:06:09.289 & $150\times111$, -24    & 5.48               & 0.064              & $36.3  \pm 0.1$   & $ 8.40 \pm 0.84$ \\
    MHO\,6  & 04:32:22.128 & $+$18:27:42.286 & $104\times72$,   35    & 4.62               & 0.025              & $48.40 \pm 0.15$  & $12.25 \pm 1.23$ \\
    J0433   & 04:33:44.670 & $+$26:15:00.080 & $128\times86$,   -8    & 6.21               & 0.024              & $37.93 \pm 0.12$  & $14.32 \pm 1.43$ \\
    CIDA\,7    & 04:42:21.022 & $+$25:20:33.996 & $101\times80$,   10    & 11.53              & 0.029              & $25.96 \pm 0.10$  & $ 6.05 \pm 0.61$ \\
    J0420   & 04:20:25.581 & $+$27:00:35.242 & $113\times82$,   22    & 6.45               & 0.021              & $17.35 \pm 0.05$  & $ 6.33 \pm 0.63$ \\
    J0415   & 04:15:58.016 & $+$27:46:16.811 & $161\times78$,  -36    & 0.44               & 0.052              & $ 0.96 \pm 0.29$  & $ 0.22 \pm 0.02$ \\
    \hline
    \hline
\end{tabular}
\end{center}
\end{table*}

\begin{table*}[ht!]
\begin{center}
\caption{$^{12}$CO and $^{13}$CO imaging summary. \label{tab:imaging_CO}}
\begin{tabular}{ |c|c|c|c|c|c|c| } 
    \hline
    \hline
    Source               & Ch. width & Detection    & Beam                 & Peak $I_\nu$    &  RMS noise & Flux  \\
                         & (km\,s$^{-1}$) & LSRK    & (mas$\times$mas,     & (mJy\,beam$^{-1}$ & (mJy\,beam$^{-1}$) & (Jy) \\
                         &           & (km\,s$^{-1}$) &  deg)              &  km\,s$^{-1}$)  &            &       \\
    \hline
    CIDA\,1  $^{12}$CO      & 0.20      &  1.7 - 11.1  & $268\times173$,  1   & 91.2            & 4.0        &  0.725 \\
\hspace{1.1cm}$^{13}$CO  & 1.00      &  2.0 - 9.1   & $118\times83$,  36   & 21.6            & 1.3        &  0.215 \\
    \hline
    MHO\,6 $^{12}$CO     & 0.25      &  0.65 - 10.1 & $115\times81$,  35   & 33.8            & 2.0        &  3.179 \\
\hspace{1.1cm}$^{13}$CO  & 0.90      &  2.0 - 9.1   & $118\times83$,  36   & 21.6            & 1.3        &  1.061 \\
    \hline
    J0433 $^{12}$CO      & 0.35      &  0.7 - 11.2  & $165\times110$, -7   & 72.7            & 2.2        &  1.241 \\
\hspace{1.2cm}$^{13}$CO  & 1.0       &  1.0- 11.0   & $159\times110$, -8   & 22.7            & 1.3        &  0.618 \\
    \hline
    CIDA\,7  $^{12}$CO      & 0.4       &  0.8 - 8.4   & $168\times148$, -1   & 61.9            & 2.6        &  1.065 \\
\hspace{1.2cm}$^{13}$CO  & 1.0       &  3.0 - 8.0   & $170\times149$,  0   & 22.8            & 1.6        &  0.160 \\
    \hline
    J0420 $^{12}$CO     & 0.45       & 7.75 - 10.9  & $135\times98$,  23   & 35.1            & 1.5        &  0.224 \\
\hspace{1.1cm}$^{13}$CO & 0.90       &  5.5 - 10.9  & $133\times96$,  22   &  8.2            & 1.0        &  0.038 \\
    \hline
    J0415 $^{12}$CO      & 0.8       &  4.4 - 8.4   & $216\times142$, -38  & 32.1            & 3.8        &  0.020 \\
\hspace{1.1cm}$^{13}$CO  & 1.0       &     ---      & $222\times141$, -40  &  ---            & 2.6        & <0.004 \\
    \hline
    \hline
\end{tabular}

\end{center}
\end{table*}

\section{Spectral profiles, eddy fitting,  and channel maps} \label{appendix:channel}

\begin{figure*}
 \centering
        \includegraphics[width=16cm]{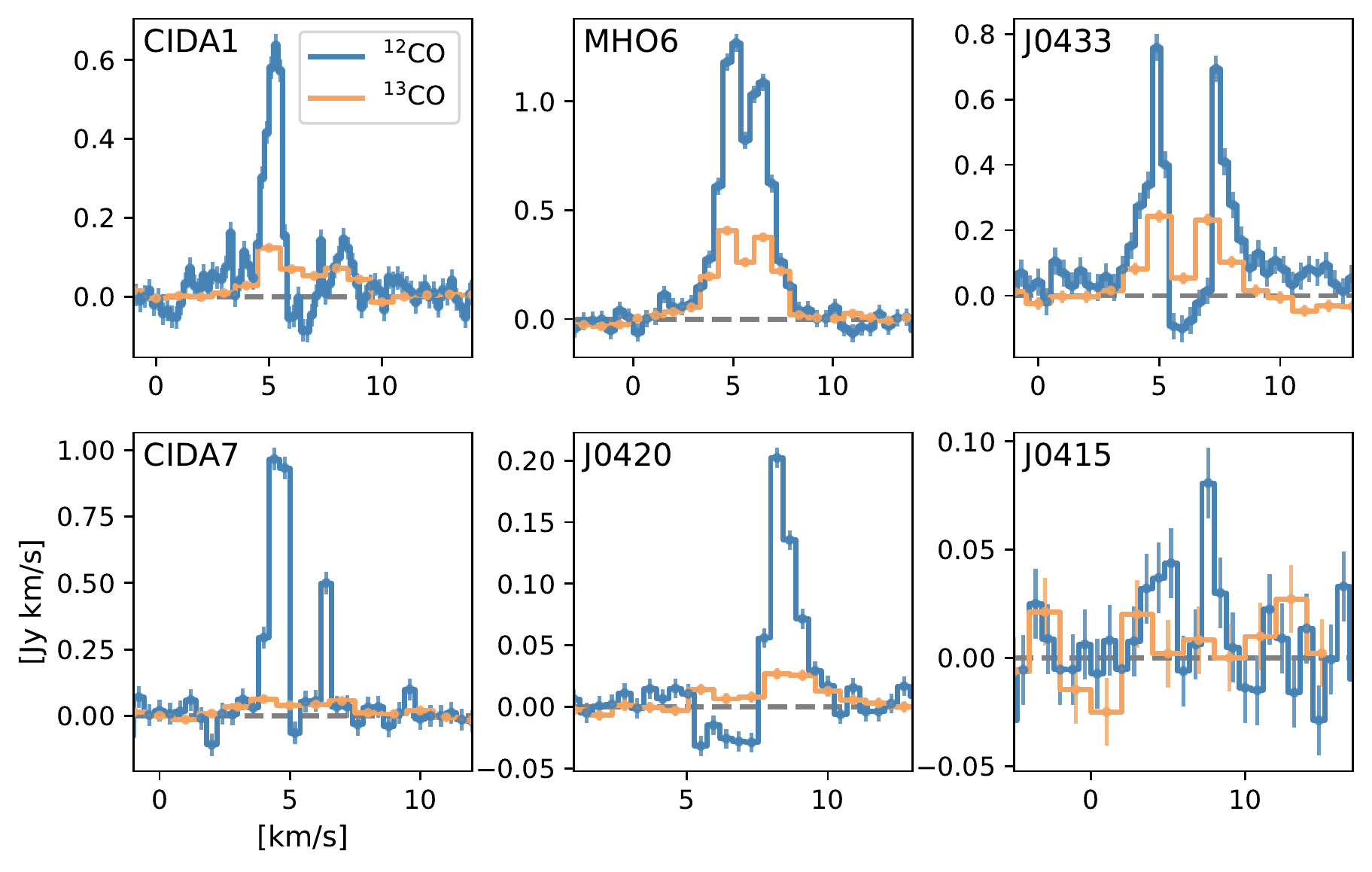}
   \caption{Spectral profile of the $^{12}$CO and $^{13}$CO in our sample. The gray line denotes the $0\,$Jy  level.}
   \label{fig:SP_CO}
\end{figure*}

The spectral profiles of $^{12}$CO and $^{13}$CO for each source, shown in Figure \ref{fig:SP_CO}, were obtained with the \texttt{CASA 5.6.2} software by placing an elliptical mask in the region where emission was detected. The errorbars represent the standard deviation of the noise, measured from the channels without emission.

\begin{figure*}
 \centering
        \includegraphics[width=16cm]{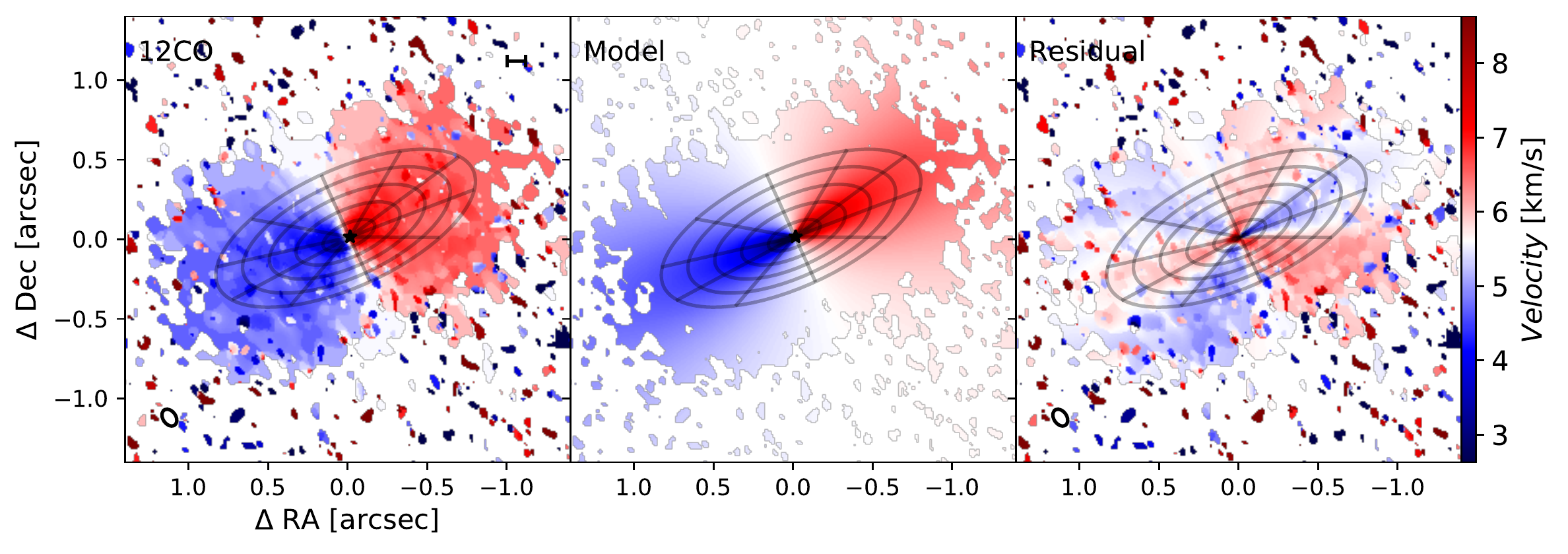} \\
        \includegraphics[width=16cm]{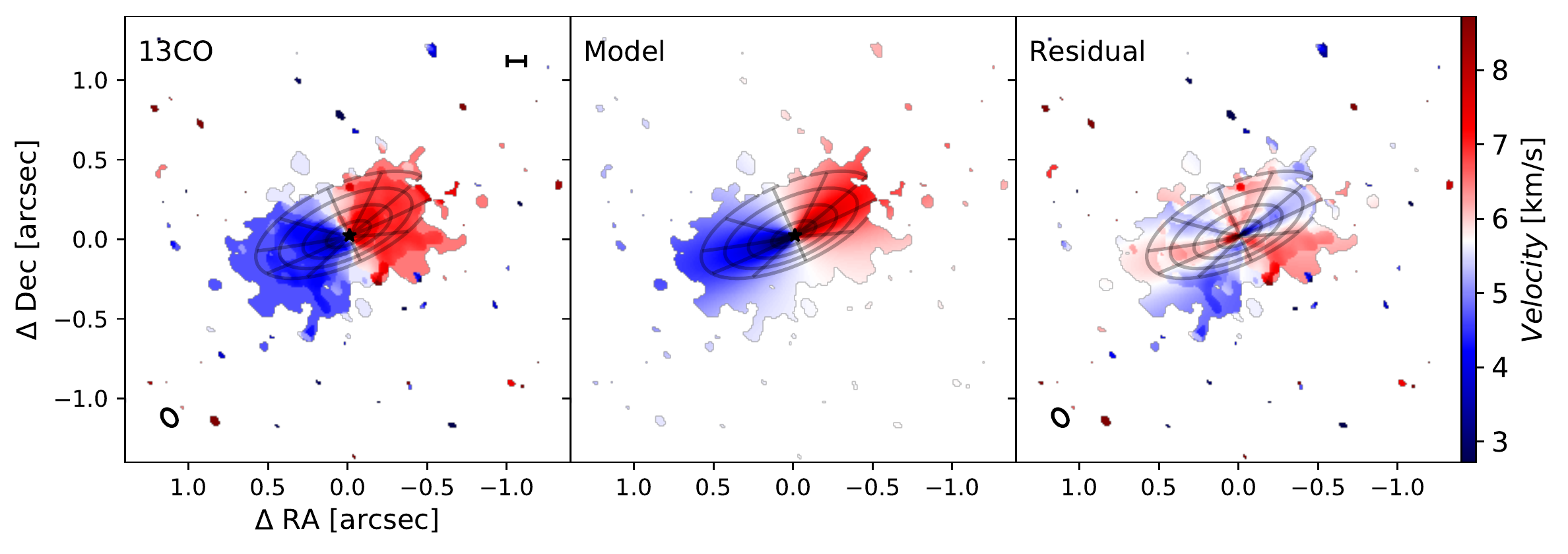} \\
   \caption{Keplerian fitting of $^{12}$CO and $^{13}$CO. The grid shows the best surface recovered, extending up to the distance of the mask for the fitting.}
   \label{fig:MHO6_COmodel}
\end{figure*}

The moment images shown in Figure \ref{fig:MHO6_COmodel} were also computed with \texttt{CASA 5.6.2} by clipping the emission in each channel at the $3\,\sigma$  level. For the Keplerian fitting with \texttt{eddy} \citep{eddy}, we fixed the inclination to be one recovered from the continuum UV-modeling. The fitting was performed inside an elliptical masked region, with the same inclination of the source, extending up to $0.9''$ in the $^{12}$CO emission, and up to $0.6''$ in the $^{13}$CO. Each image was modeled separately

The models and residuals shown in Figure \ref{fig:MHO6_COmodel} allowed the center of the source ($x_0$, $y_0$), the position angle (PA), the mass of the star ($M_\star$), the central velocity (VLSR), and the vertical structure ($Z(r)=z_0\cdot r^\phi$) to vary. The central velocity was consistent between the two lines, and PA was also consistent with continuum. For the stellar mass, $^{12}$CO would give a value of $M_\star = 0.195\pm0.02\,M_\odot$, and $M_\star = 0.19\pm0.1\,M_\odot$  from $^{13}$CO. Fixing some of these parameters to simplify the model, or using images obtained with \texttt{bettermoments}, would change $M_\star$ between $0.16\sim0.24\,M_\odot$, but the structure of residuals would remain the same.

\begin{figure}
 \centering
        \includegraphics[width=8cm]{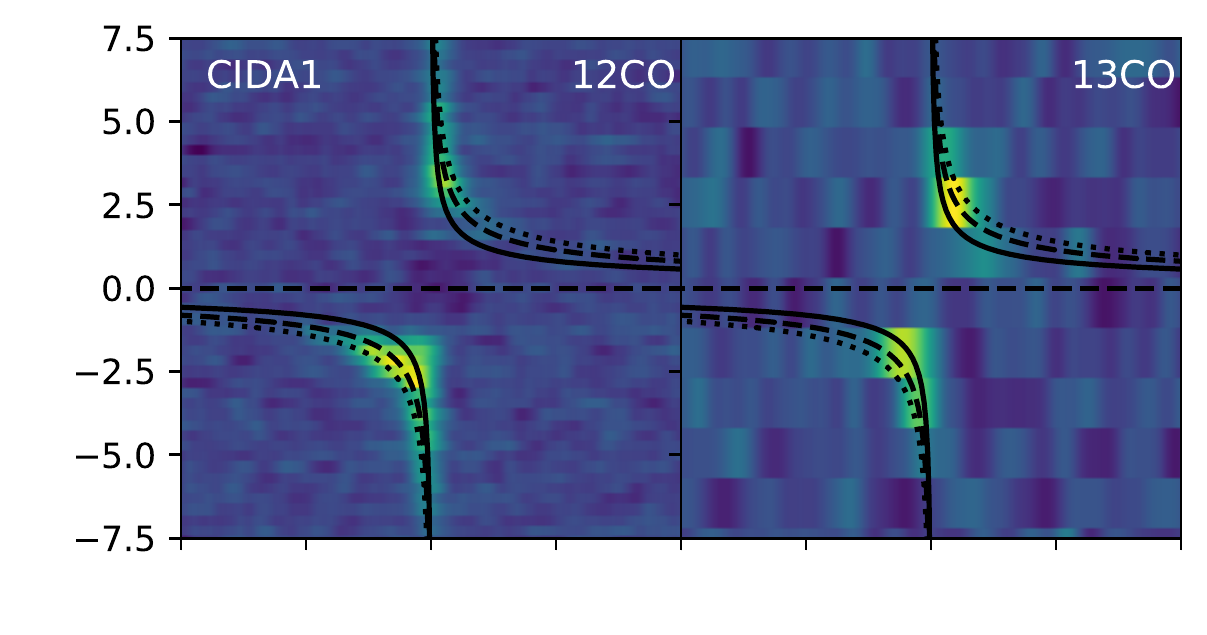} \\
    \includegraphics[width=8cm]{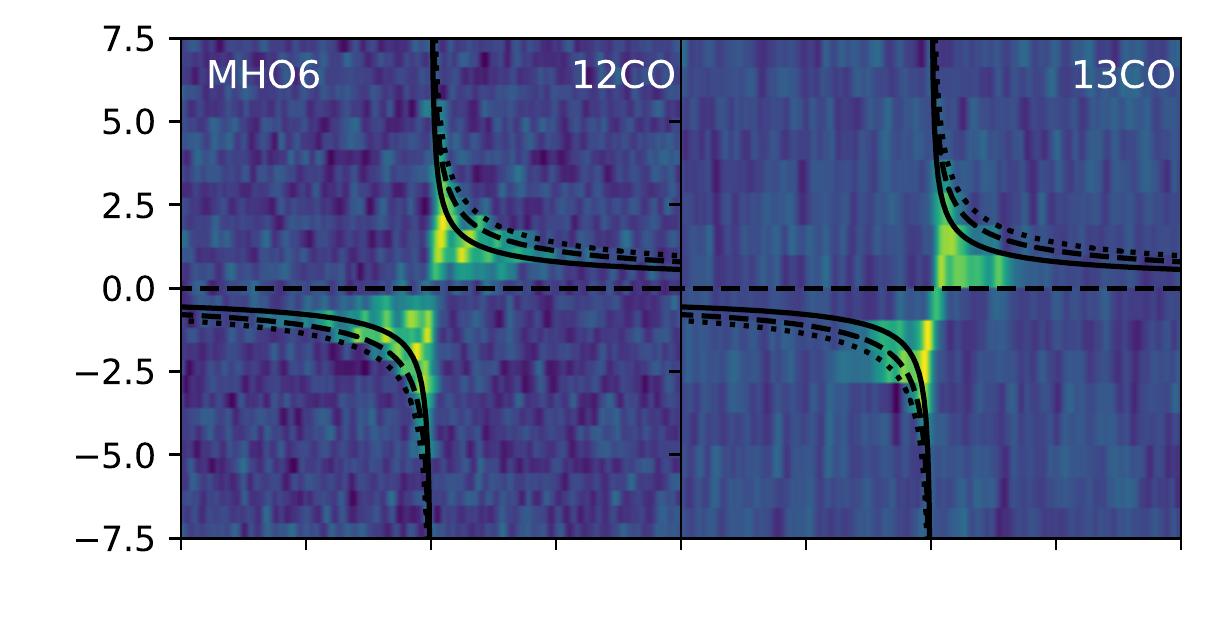} \\
    \includegraphics[width=8cm]{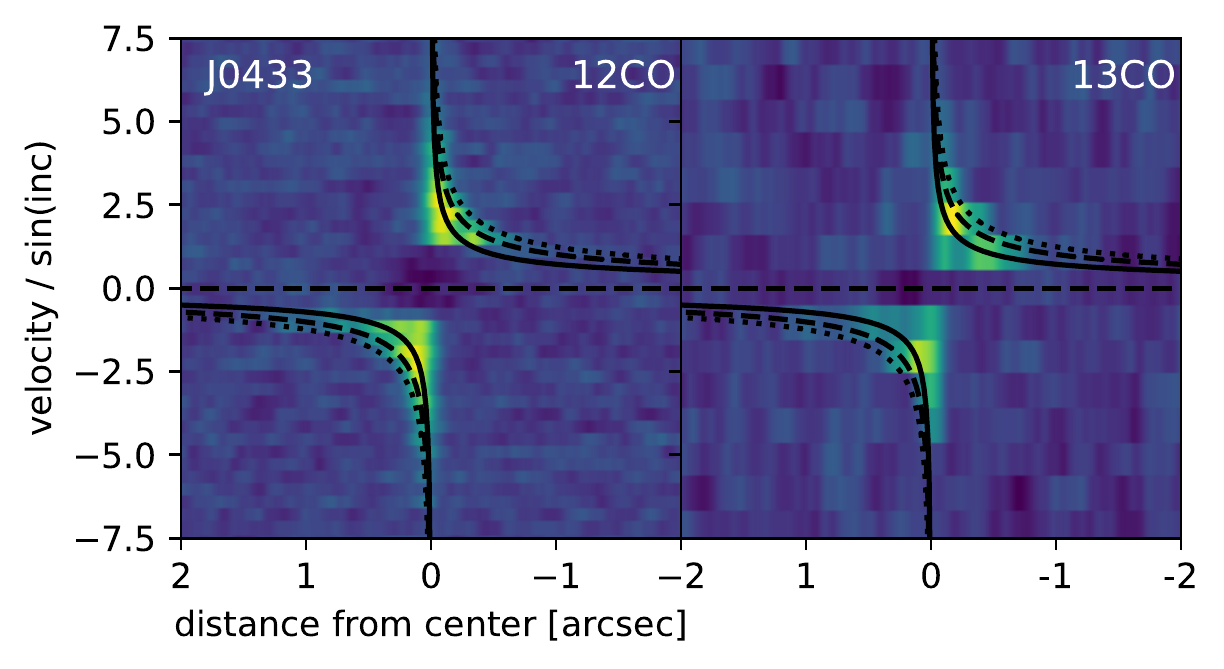} \\
        \caption{Position velocity diagrams for the sources where substructures are detected in the continuum emission. Lines follow the Keplerian velocities for central mass objects of $0.1$, $0.2$, and $0.3\,M_\odot$  (solid, dashed, and dotted, respectively).}
   \label{fig:PV_subs}
\end{figure}

\begin{figure}
 \centering
        \includegraphics[width=8cm]{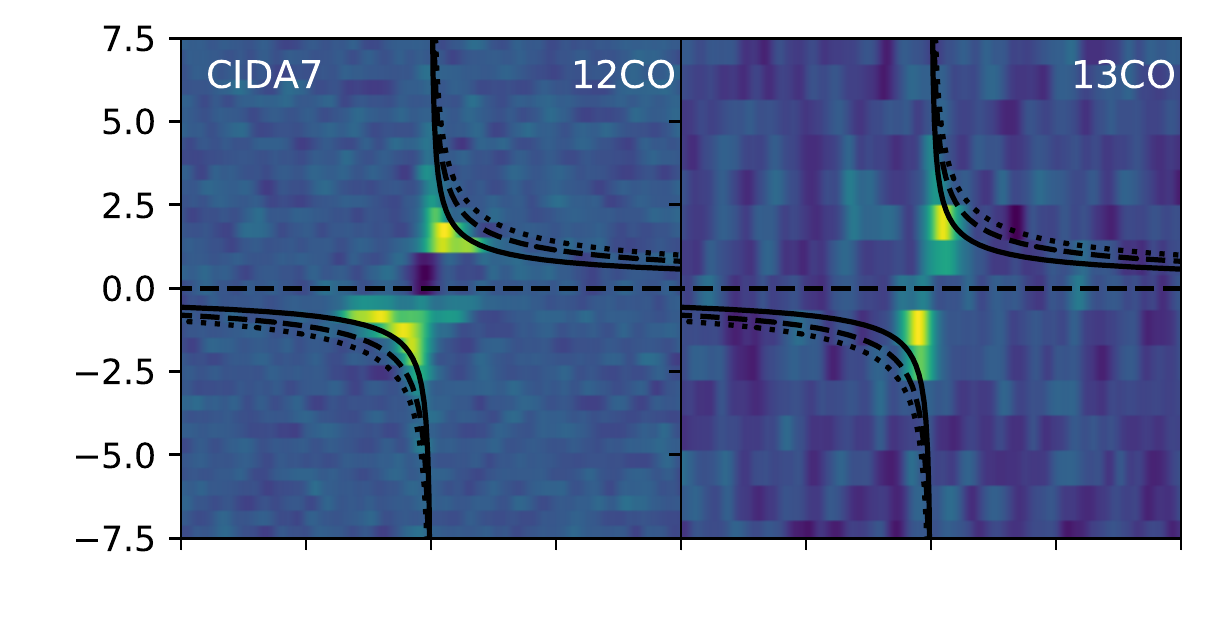} \\
    \includegraphics[width=8cm]{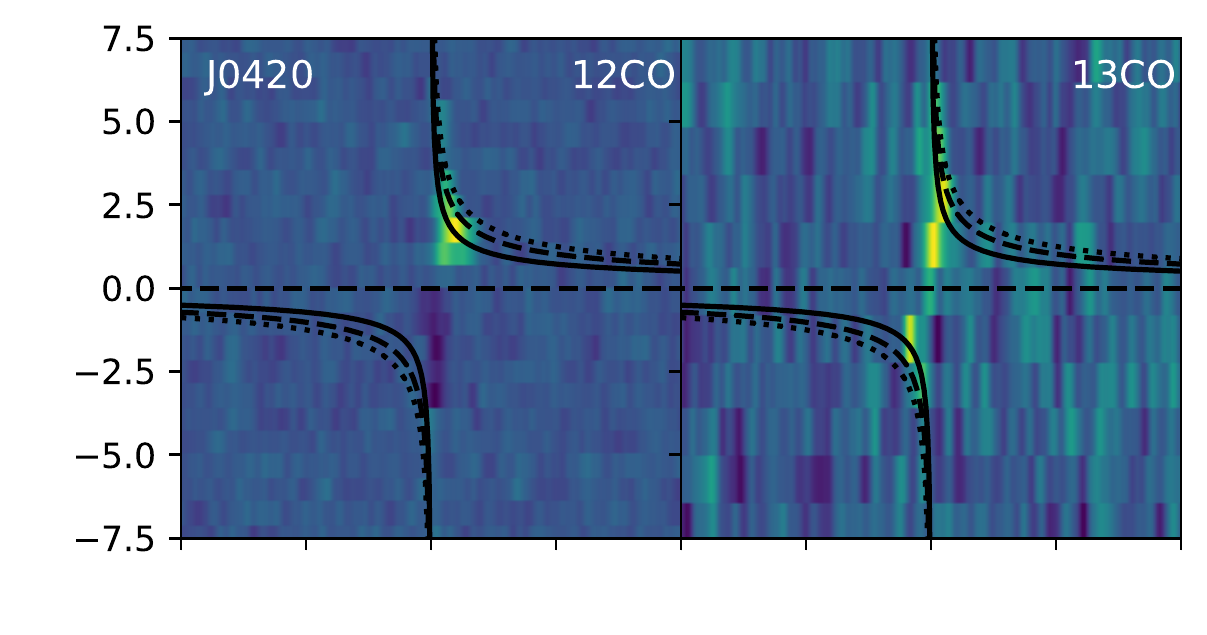} \\
    \includegraphics[width=8cm]{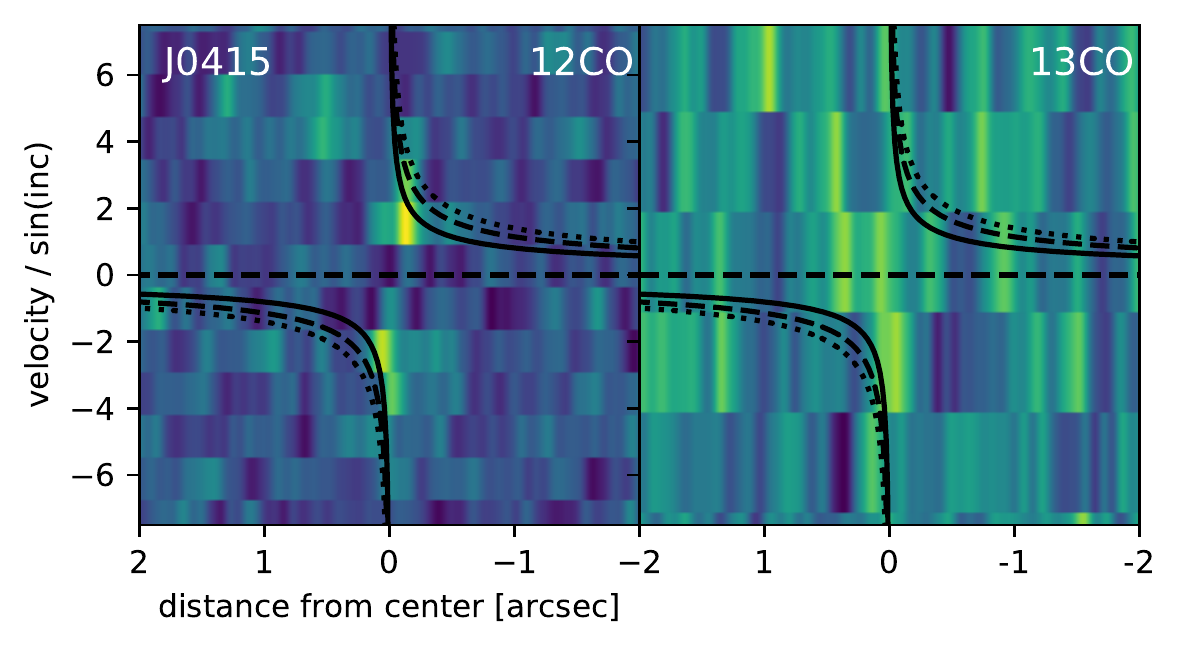} \\
        \caption{Same as Figure \ref{fig:PV_subs}, but for the sources without a substructure.}
   \label{fig:PV_nosubs}
\end{figure}

\begin{figure*}
 \centering
        \includegraphics[width=17cm]{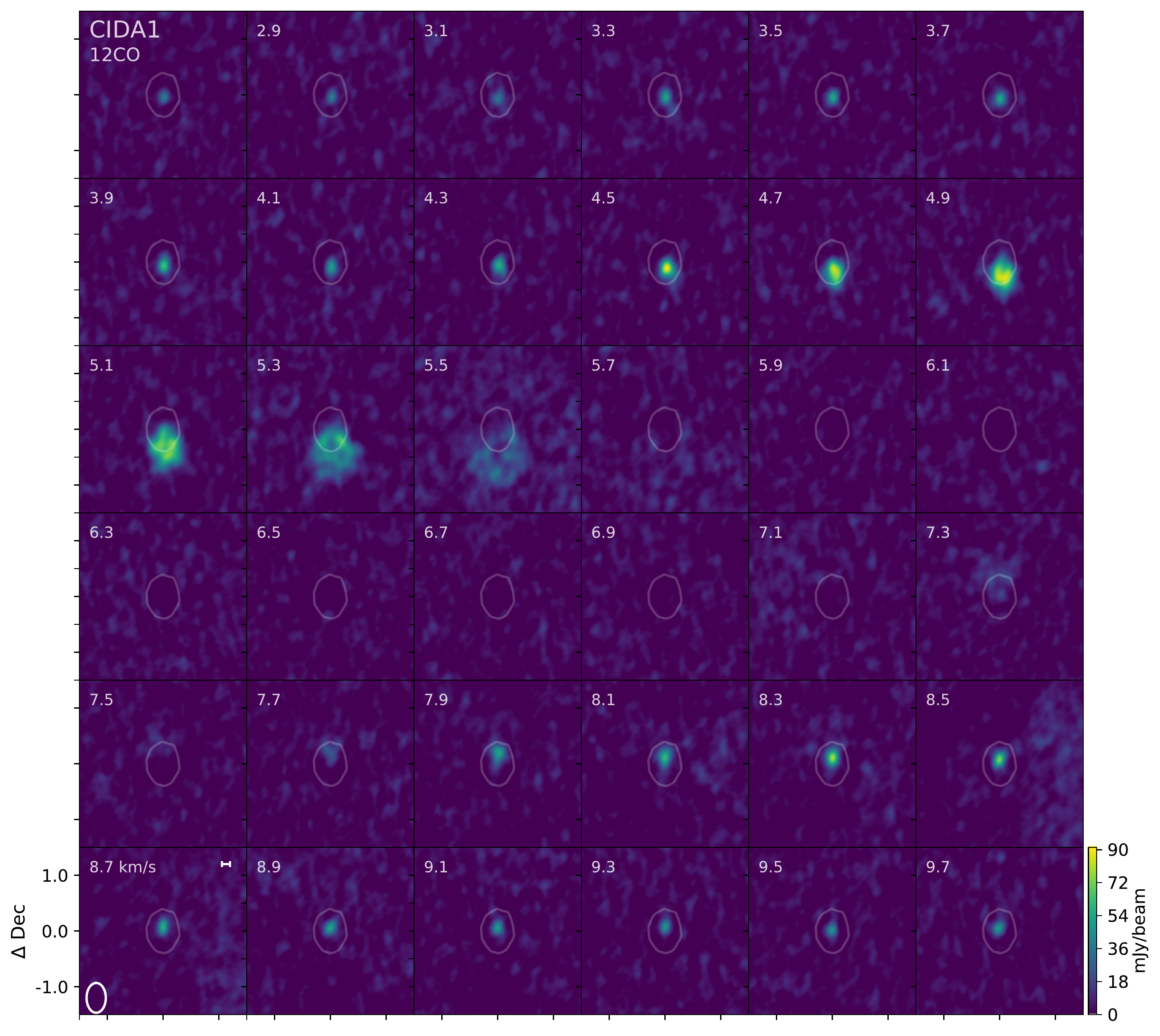}
        \includegraphics[width=17cm]{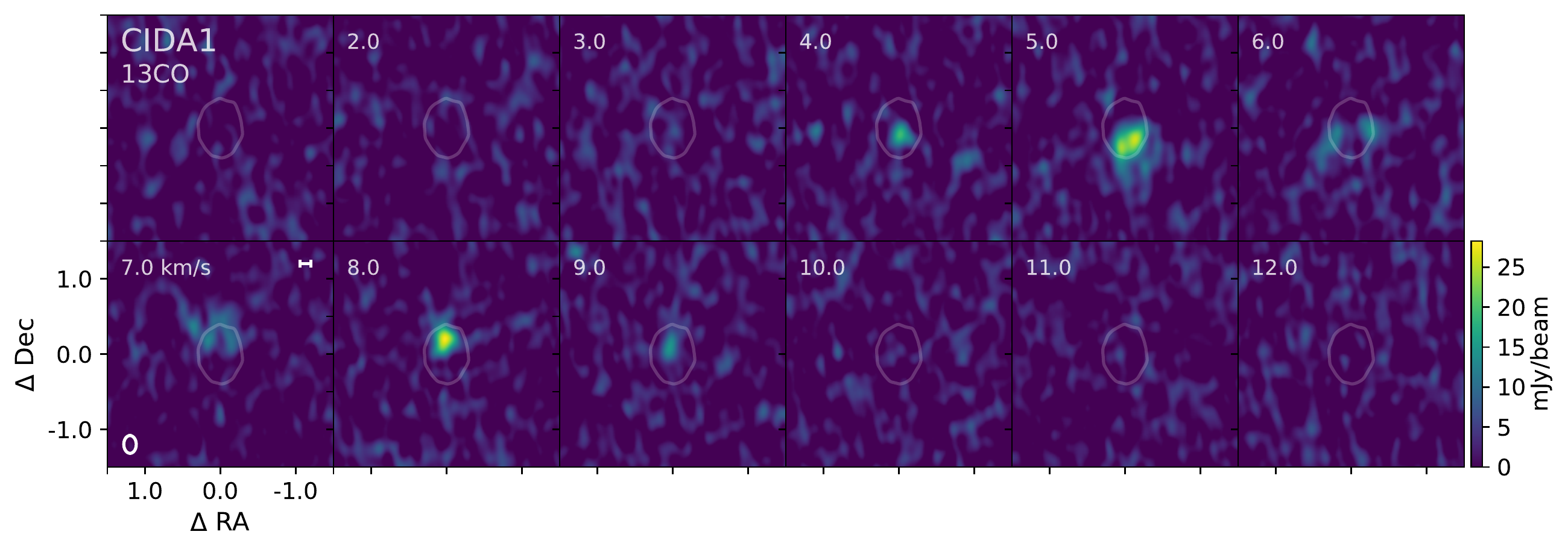}
   \caption{Images of CIDA\,1  $^{12}$CO channel maps are shown in the top, and $^{13}$CO channel maps are shown in the bottom. Each square is 3.0'' in width and height, centered at the same position as the dust continuum image. The contour level traces the $5\,\sigma$  emission in the continuum image. The scale bar in the lower left panel is $20\,$au in size, and the beam size is found in the lower left corner of the same panel. Central velocities in km/s of each channel are given in the upper left corner of each panel.}
   \label{fig:chanmap_cida1}
\end{figure*}

\begin{figure*}
 \centering
        \includegraphics[width=17cm]{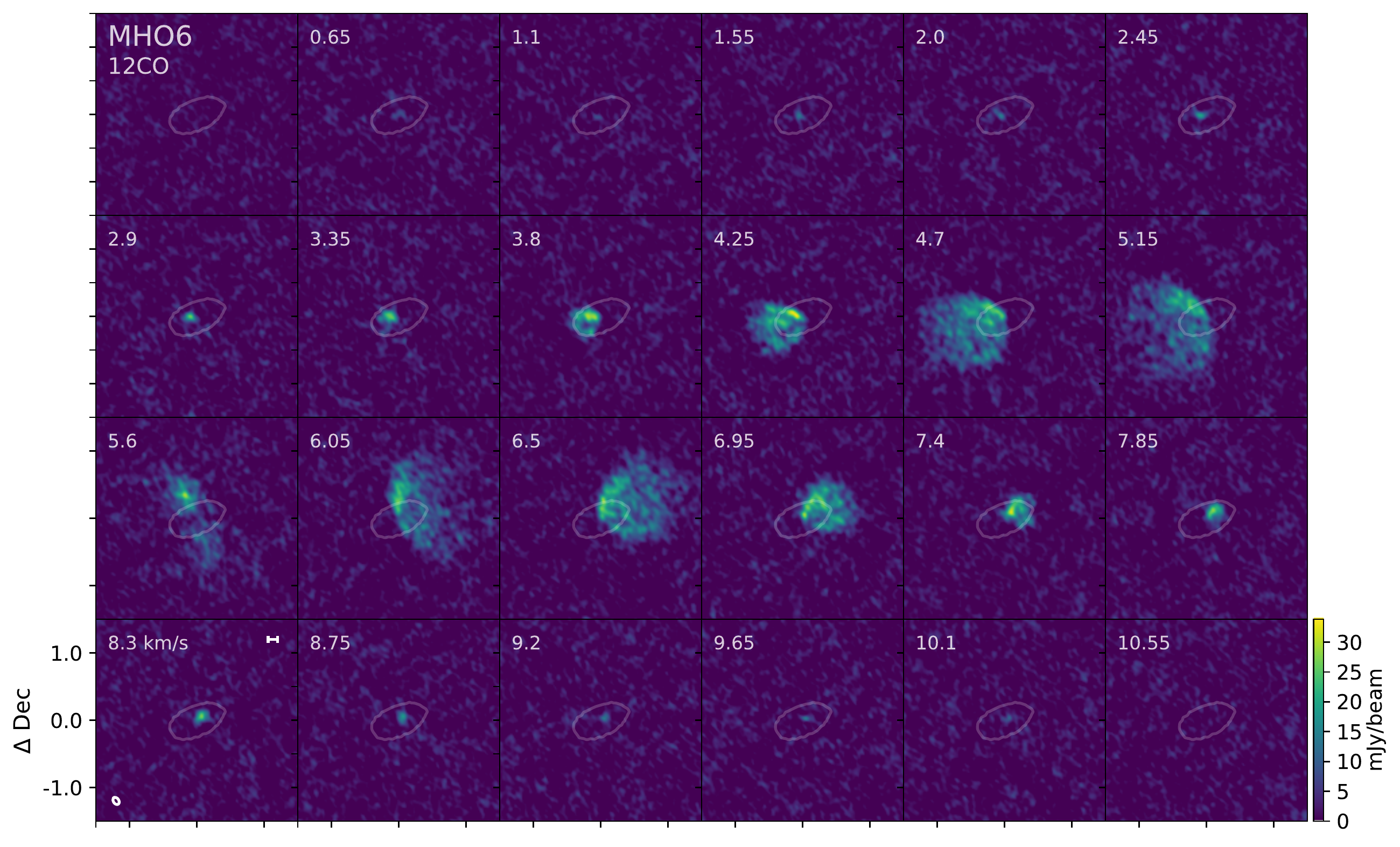}
        \includegraphics[width=17cm]{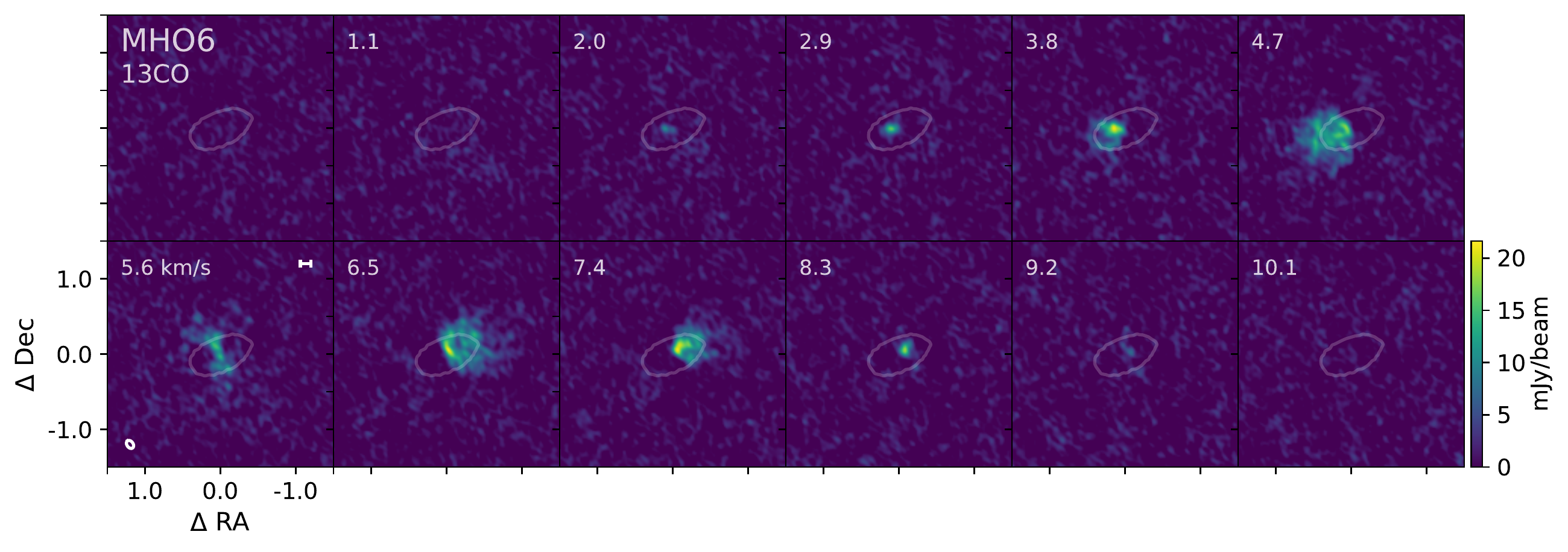}
   \caption{Same as Figure \ref{fig:chanmap_cida1}, but for MHO\,6. }
   \label{fig:chanmap_mho6}
\end{figure*}

\begin{figure*}
 \centering
        \includegraphics[width=17cm]{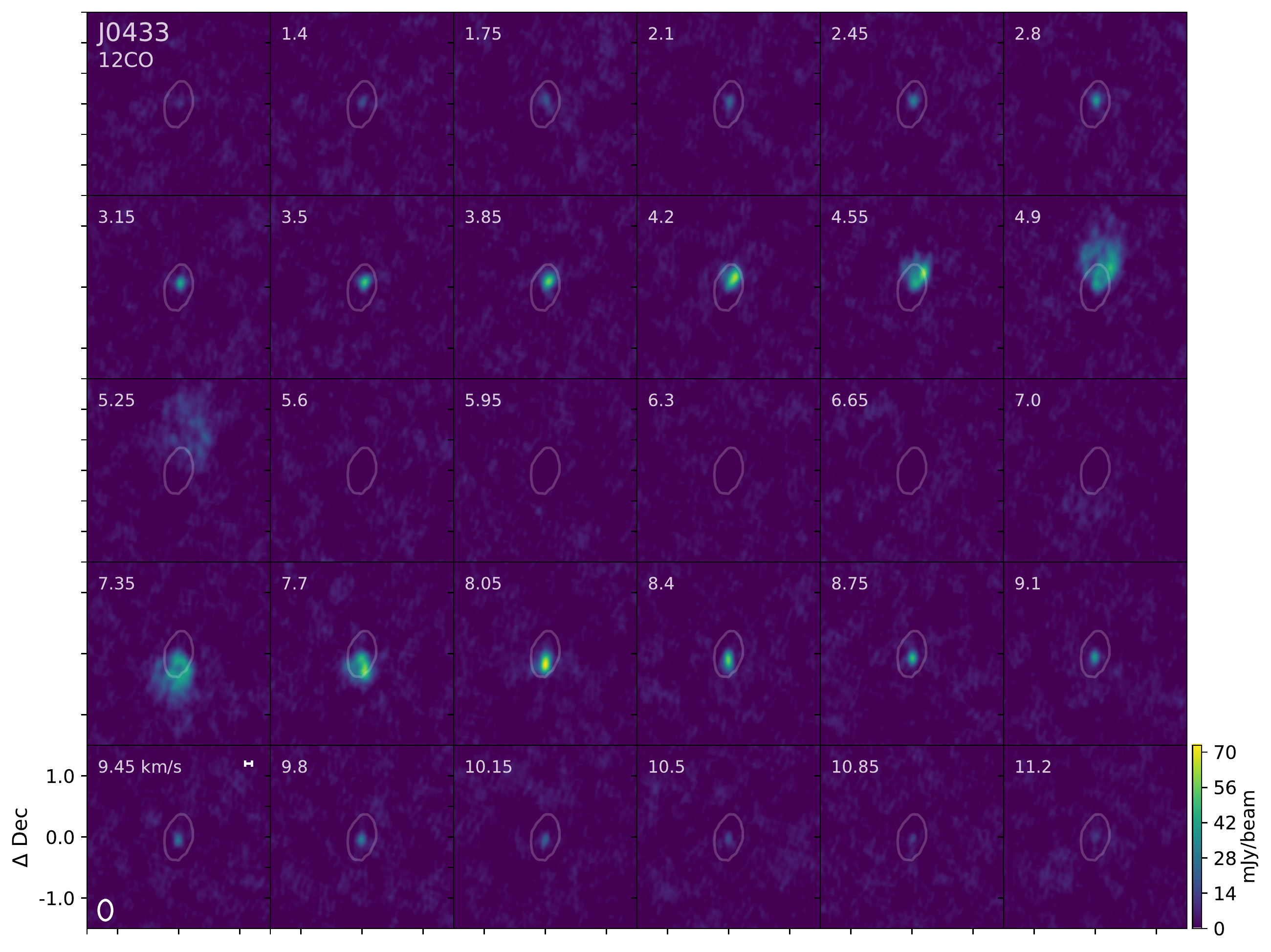}
        \includegraphics[width=17cm]{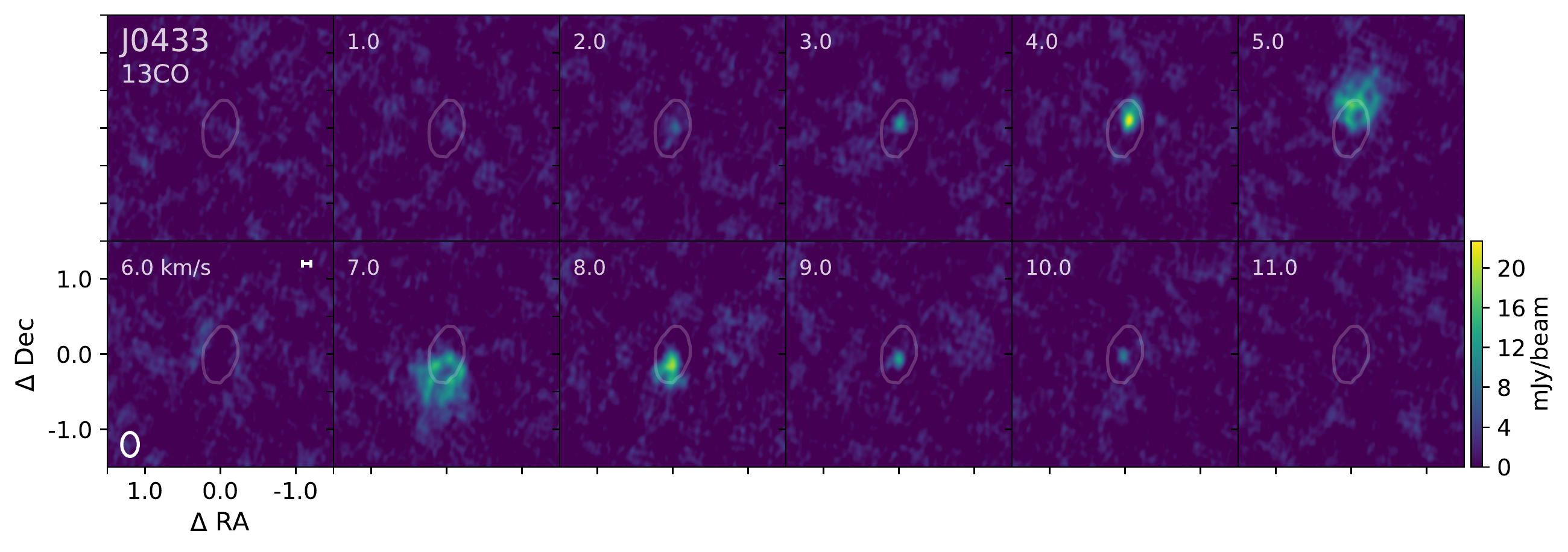}
   \caption{Same as Figure \ref{fig:chanmap_cida1}, but for J0433.}
   \label{fig:chanmap_j0433}
\end{figure*}

\begin{figure*}
 \centering
        \includegraphics[width=17cm]{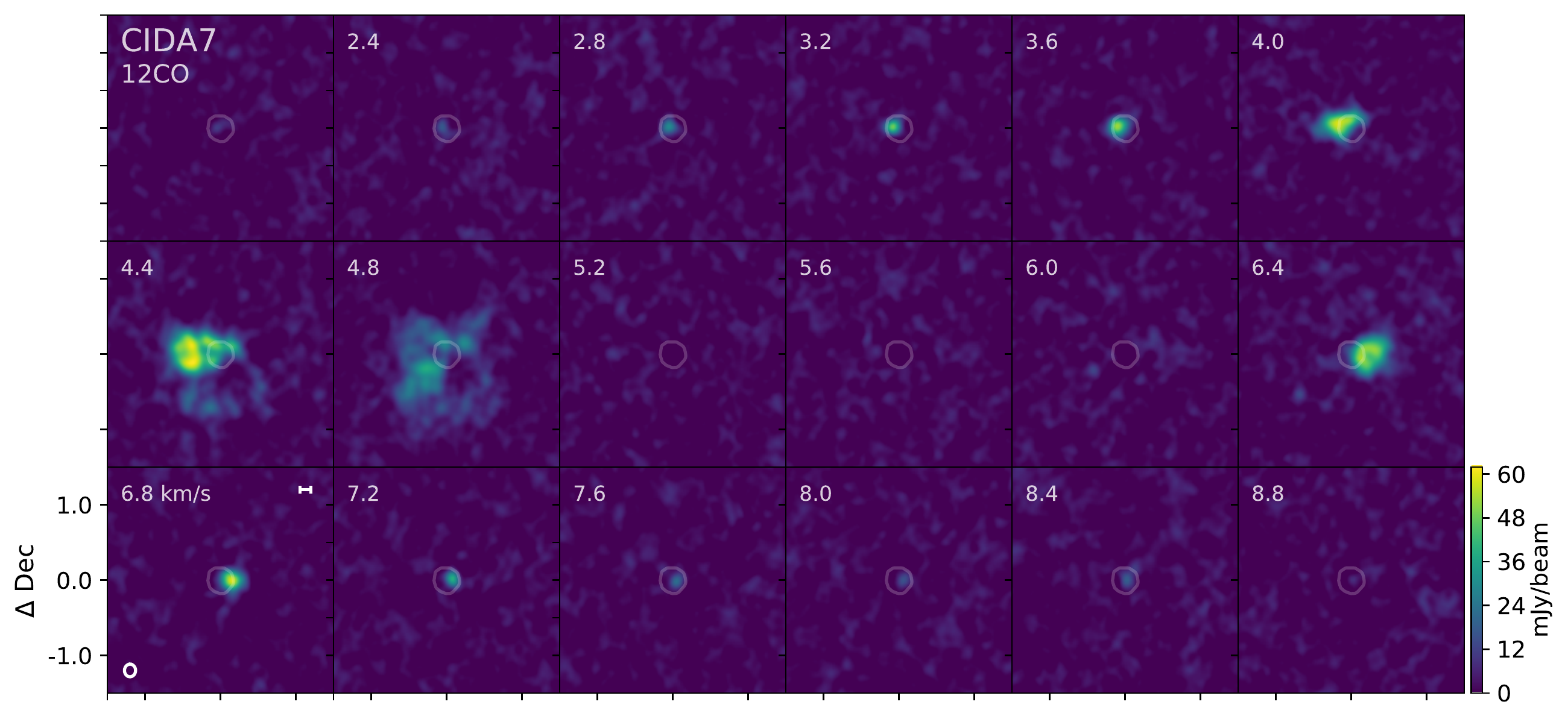}
        \includegraphics[width=17cm]{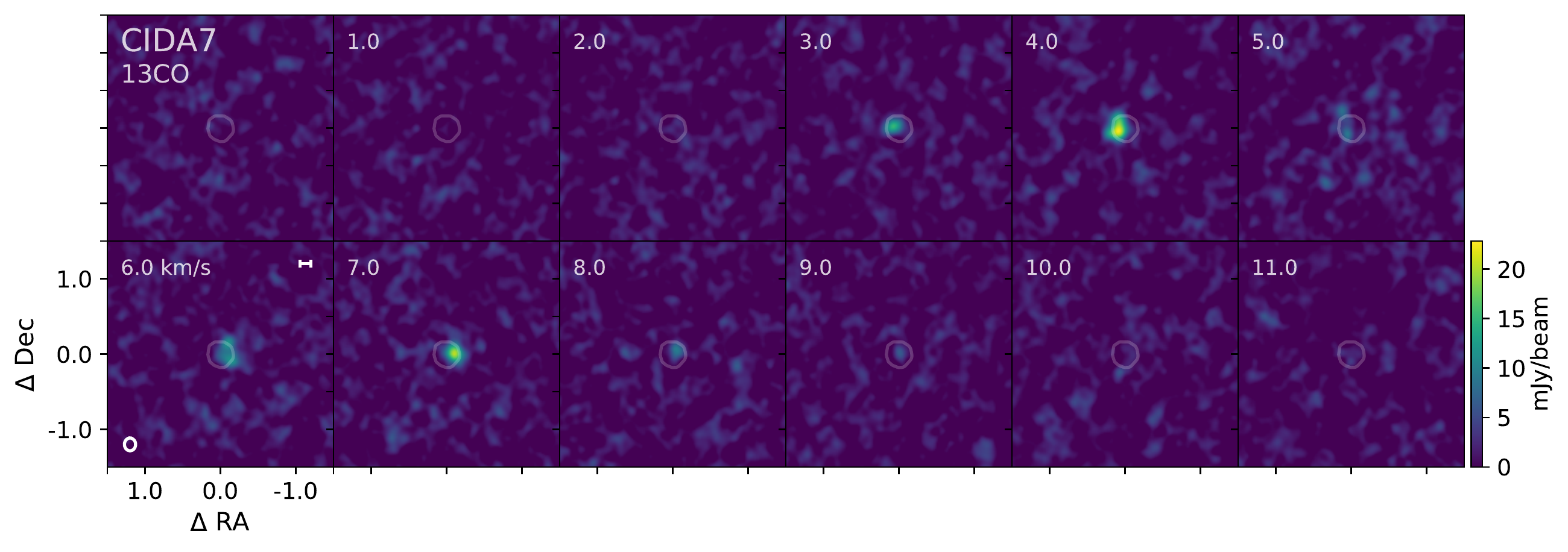}
   \caption{Same as Figure \ref{fig:chanmap_cida1}, but for CIDA\,7. }
   \label{fig:chanmap_cida7}
\end{figure*}

\begin{figure*}
 \centering
        \includegraphics[width=17cm]{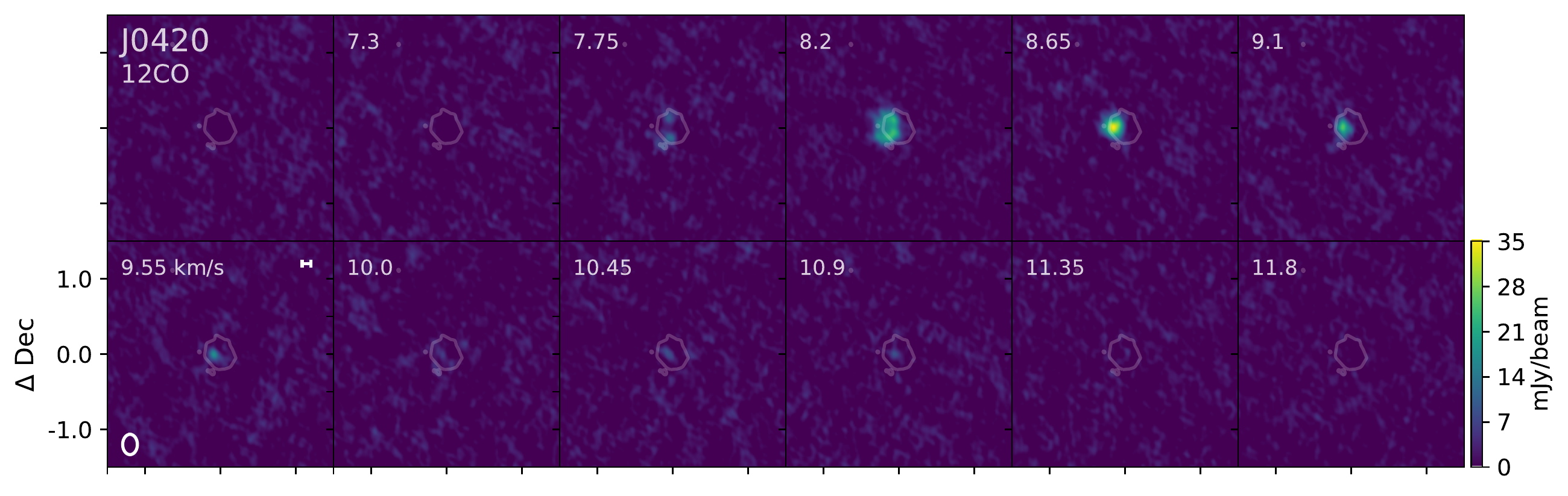}
        \includegraphics[width=17cm]{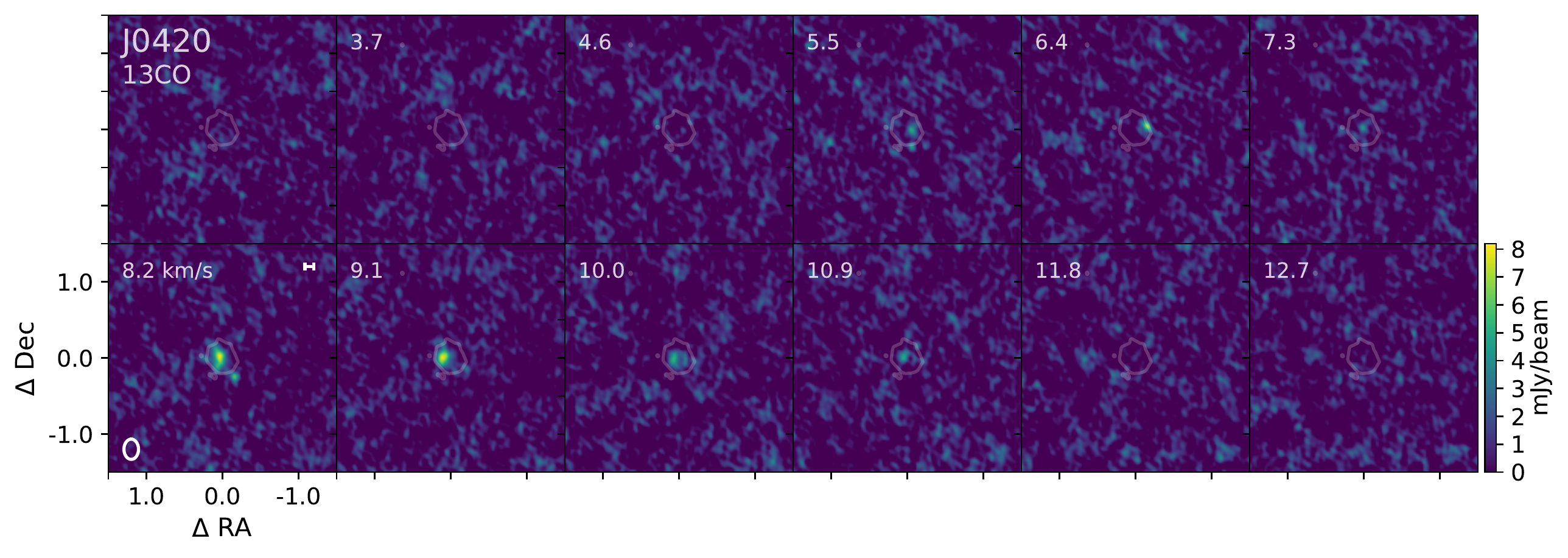}
   \caption{Same as Figure \ref{fig:chanmap_cida1}, but for J0420.}
   \label{fig:chanmap_j0420}
\end{figure*}

\begin{figure*}
 \centering
        \includegraphics[width=17cm]{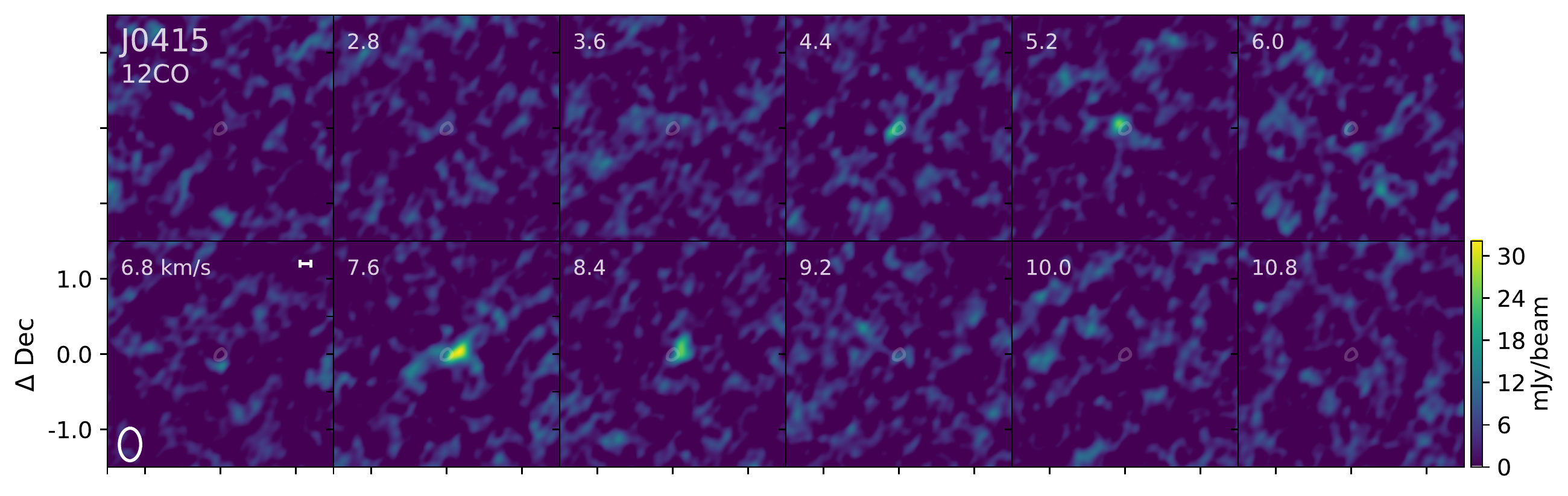}
        \includegraphics[width=17cm]{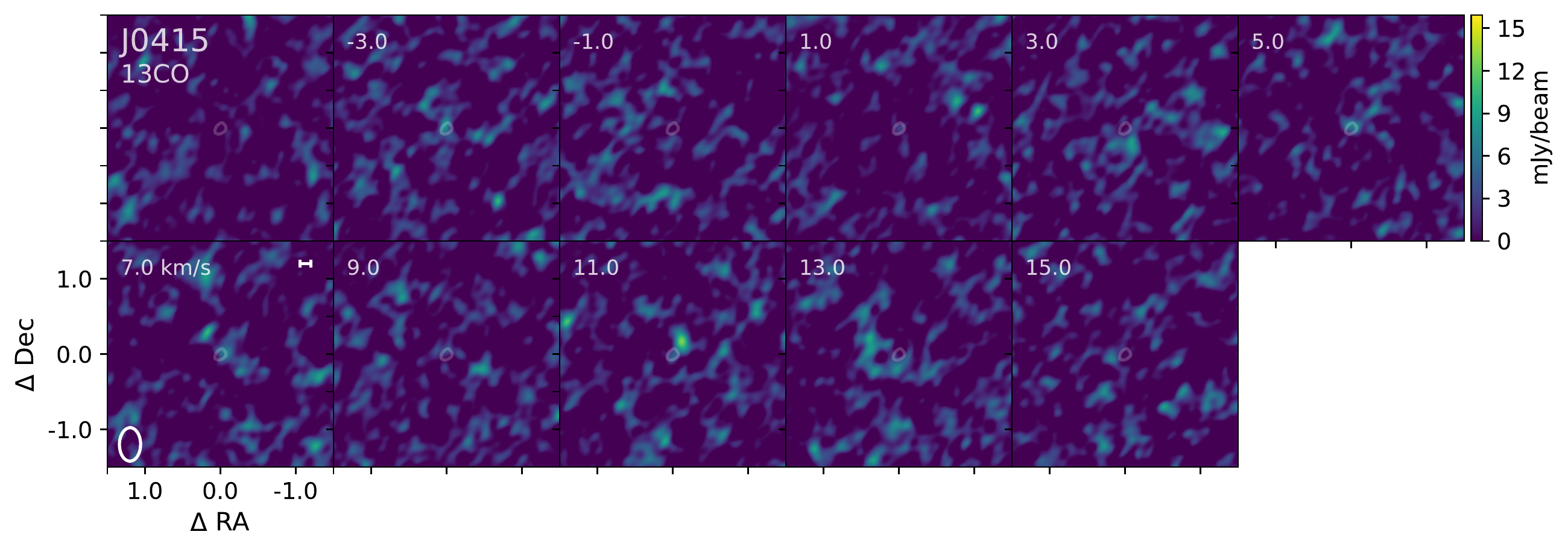}
   \caption{Same as Figure \ref{fig:chanmap_cida1}, but for J0415.}
   \label{fig:chanmap_j0415}
\end{figure*}

\end{appendix}

\end{document}